\documentclass[a4paper,11pt]{article}
\pdfoutput=1 
\usepackage{jcappub} 
\usepackage[T1]{fontenc}
\usepackage{subcaption}
\usepackage{hyperref}
\usepackage[utf8]{inputenc}
\begin{document}

\title{Angular power spectrum analysis on current and future high-energy neutrino data}

\author[a]{Ariane Dekker}
\author[a,b]{and Shin'ichiro Ando}

\affiliation[a]{GRAPPA Institute, University of Amsterdam, 1098 XH Amsterdam, Netherlands}
\affiliation[b]{Kavli Institute for the Physics and Mathematics of the Universe, University of Tokyo, Kashiwa, Chiba 277-8583, Japan}

\emailAdd{a.h.dekker@uva.nl}
\emailAdd{s.ando@uva.nl}

\abstract{Astrophysical neutrino events have been measured in the last couple of years, which show an isotropic distribution, and the current discussion is their astrophysical origin. We use both isotropic and anisotropic components of the diffuse neutrino data to constrain the contribution of a broad number of extra-galactic source populations to the observed neutrino sky. We simulate up-going muon neutrino events by applying statistical distributions for the flux of extragalactic sources, and by Monte Carlo method we exploit the simulation for current and future IceCube, IceCube-Gen2 and KM3NeT exposures. We aim at constraining source populations by studying their angular patterns, for which we assess the angular power spectrum. 
We leave the characteristic number of sources ($N_{\star}$) as a free parameter, which is roughly the number of neutrino sources over which the measured intensity is divided. With existing two-year IceCube data, we can already constrain very rare, bright sources with $N_{\star}\lesssim$100. This can be improved to $N_{\star}\lesssim 10^4$--$10^5$ with IceCube-Gen2 and KM3NeT with ten-year exposure, constraining the contribution of BL Lacs ($N_{\star}=6\times10^{2}$). On the other hand, we can constrain weak sources with large number densities, like starburst galaxies ($N_{\star} = 10^{7}$), if we measure an anisotropic neutrino sky with future observations. }

\maketitle
\flushbottom

\section{Introduction}
\label{sec:intro}

High-energy neutrinos with energies between tens of TeV and multiple PeV have been detected with IceCube at South Pole~\cite{IceCubeScience,Aartsen:2014gkd}.
Distribution of these neutrino events are isotropic, suggesting towards dominant contributions from extragalactic sources~\cite{Aartsen:2016xlq,IceCubeICRC}. Moreover, the observed neutrino flux is consistent with the extragalactic gamma-ray flux and the ultrahigh-energy cosmic ray flux, which might indicate towards the same population of sources~\cite{Ahlers:2018fkn,Waxman:2013zda}. 
Theoretical models predict the production of neutrinos within or in the surroundings of cosmic ray accelerator through the decay of charged pions, whereby the pions are produced by hadronuclear collision~\cite{Loeb:2006tw,Murase:2008yt,Fang:2017zjf,Murase:2013rfa} or photo-hadronic interactions~\cite{Murase:2015xka,Rodrigues:2017fmu,Guepin:2017dfi} with cosmic rays. Gamma-rays are produced alongside through the decay and production of neutral pions. 
The two channels result in different ratios between gamma-ray and neutrino flux and combining these observations will therefore offer insight in the production mechanism, as well as an hint for possible neutrino candidate sources. The Fermi-LAT data showed that the gamma-ray point-source emission is dominated by blazars~\cite{Lisanti:2016jub,Acero:2015hja,Ackermann:2015uya}. Indeed, very recently, temporal and positional associations of IceCube events with a blazar, TXS 0506+056, has been found at $\sim$3--3.5$\sigma$ level~\cite{IceCube:2018cha,IceCube:2018dnn}.
Even though these could still be caused by atmospheric neutrino background, if confirmed, the blazars will be established as one of the contributing sources of high-energy neutrinos through cosmic ray acceleration followed by hadronic interactions.

However, it is possible that more than one source population gives significant contributions to the measured flux of the high-energy neutrinos.
In fact, Ref.~\cite{Murase:2018iyl} pointed out that the blazar flares as was identified by Refs.~\cite{IceCube:2018cha,IceCube:2018dnn} were responsible up to $\sim$10\% --- only a sub-dominant fraction of the total intensity of the IceCube neutrinos~\cite{IceCubeScience,Aartsen:2014gkd}. Therefore, alternative sources are being studied for temporal and positional coincidence with neutrino signals, whereby muon track events are favorable probes due to their good angular resolution and large statistics~\cite{IceCubePS}. 
Nevertheless, besides XS 0506+056 no other point-source was found, which sets constraints on the contribution of source populations, e.g., gamma-ray bursts~\cite{Aartsen:2018fpd}, core-collapse supernovae~\cite{Esmaili:2018}, active galactic nuclei~\cite{Murase:2015ndr}, starburst galaxies~\cite{Bechtol:2015uqb} and galaxy clusters and groups~\cite{Zandanel:2015vva}. Moreover, also a galactic contribution is predicted, which is found to be relatively small and we will thus not consider this in our analysis~\cite{Aartsen:2017ujz,Halzen:2016seh}. 

Another method in order to reveal sources that give dominant contribution to the measured diffuse flux, is by studying angular clustering of registered events~\cite{IceCubeAnis,MuraseWaxman,Ando:2017xcb} and is shown to be more efficient than the conventional point-source searches~\cite{IceCubePS} for individually bright, rare source populations such as blazars~\cite{Ando:2017xcb}. References~\cite{Mertsch:2016hcd,MuraseWaxman} set, for instance, stringent limits on the number density of neutrino sources, which has the advantage that a wide range of neutrino source populations is discussed without using prior theoretical models. As complementary method, one-point fluctuation analysis~\cite{Feyereisen2016} maximises information obtained in single pixels, which helps to constrain both extragalactic and Galactic source contributions~\cite{Feyereisen:2017fnk}.

In this work, we study the angular clustering of current and future up-going muon neutrino events, observed by IceCube, IceCube-Gen2 and KM3NeT, by performing an angular power spectrum analysis. Instead of using an equal flux for all sources, we apply statistical distributions based on cosmological considerations, to assign a flux to each extragalactic neutrino source. We use observables as the intensity and isotropy that were measured by IceCube~\cite{Aartsen:2015rwa}. By performing Monte Carlo simulations, we test the sensitivity of this analysis to a broad range of potential source classes.
The work extends fully analytic treatment of Ref.~\cite{Ando:2017xcb} by adopting more realistic simulations of the expected sky map of neutrino events of given exposure.
By repeating these Monte Carlo simulations, we make error estimates more robust by taking cross correlations between different multipole bins fully into account.
This turns out to yield a significant effect on projections, which weakens sensitivity with the angular power spectrum approach compared with the earlier estimates of Ref.~\cite{Ando:2017xcb} based on analytic calculations.
We use the characteristic number of sources ($N_{\star}$) to assess the sensitivities, which is roughly the number of neutrino sources over which the measured intensity is divided, and bright sources like BL Lacs have therefore smaller values ($N_{\star}=6\times10^{2}$) than numerous, weak sources like starburst galaxies ($N_{\star} = 10^{7}$). We are still able to already test the case of a characteristic number of sources of $\lesssim$100 with the existing two-year IceCube data, and this can be improved to $\lesssim 10^4$--$10^5$ with IceCube-Gen2 and KM3NeT with ten-year exposure.

The paper is organized as follows. In Sec.~\ref{sec:1} we discuss the data sample used in our analysis, consisting of 21 high-energy up-going muon events from 2 years of IceCube exposure. Furthermore, we discuss the full-sky map simulation, where we applied statistical distributions for the astrophysical flux and the flux model from Ref.~\cite{Aartsen:2015rwa} for the atmospheric flux. In Sec.~\ref{sec:2} we discuss the angular power spectrum analysis. In Sec.~\ref{sec:3} we show the results using the data sample, where we can already constrain very bright and rare sources, and the results using future neutrino data. Subsequently, we discuss the implications of the result with respect to known source populations in Sec.~\ref{sec:4}, and we find significant constraints for sources as BL Lacs and starburst galaxies. Finally, we conclude in Sec.~\ref{sec:5}.

\section{Half-sky map of neutrino events}\label{sec:1}
In the following, we discuss the data sample and the sky map of up-going high-energy muon neutrinos, detectable by current and future neutrino telescopes IceCube, IceCube-Gen2 and KM3NeT. Muon neutrinos have a good angular resolution due to the long tracks in the detectors~\cite{IceCubePS}. Moreover, up-going events traverse the Earth and thereby shield off background events from cosmic ray muons, allowing the interaction vertex to happen outside the detector and consequently enlarging the effective area~\cite{Aartsen:2016xlq}. 

\subsection{Data sample}
We obtain the data sample from two years of IceCube exposure, in which 35000 neutrino events were detected from the Northern hemisphere for $E_{\mu}\gtrsim 10^{2}$ GeV~\cite{Aartsen:2015rwa}. Most of these events are atmospheric neutrinos, produced when cosmic rays interact with the Earth's atmosphere, and only the events with the highest energies are most probable of extra-terrestrial origin. For this reason, a threshold on the energy proxy of the events is set with $E_{\mu} > 5\times 10^{4}$~GeV in the case of IceCube, resulting in 21 muon detections in the two year dataset. The high-energy upgoing muon events are downloaded from \url{https://icecube.wisc.edu/science/data/HE_NuMu_diffuse}. The energy spectra of the astrophysical component is taken to be the best-fit to the upgoing muon neutrino data~\cite{Aartsen:2015rwa}:
\begin{equation}\label{eq:spectra}
\frac{d\Phi_\nu}{dE_\nu} = 1.7\times 10^{-18}~\mathrm{GeV^{-1}~cm^{-2}~s^{-1}~sr^{-1}}\left(\frac{E_\nu}{100~{\rm TeV}}\right)^{-2.2},
\end{equation}
found by assuming an equal flavor composition and equal neutrino and antineutrino ratio. Furthermore, we need a description for the background, which are mainly conventional atmospheric neutrinos produced in the decay of pions and kaons~\cite{Gaisser:1990vg}. We use the flux model taken from the above URL, showing a much softer spectrum.

Although more recent through-going muon data sets exist, only the first two-year data can be downloaded from the web page above together with the effective area data.
For example, Ref.~\cite{Aartsen:2016xlq} discusses six-years of through-going data, but proper exposure information specific to this data set is unavailable.
Since the data sample of two years consists of only 21 events above the energy threshold of $E_{\mu} > 5\times 10^{4}$~GeV we do not apply a further energy cut. Consequently, we will have more background contamination.

\subsection{Astrophysical flux distribution}
We assume that the source flux distribution of neutrino sources is described by a broken power-law:
\begin{equation}\label{eq:1}
\frac{dN_{s}}{dF} = N_\star \times
\begin{cases} 
\left(\frac{F}{F_{\star}}\right)^{-\alpha},  & \quad F_{\star}<F,\\ 
\left(\frac{F}{F_{\star}}\right)^{-\beta},  & \quad F_{0}<F<F_{\star}\,,
\end{cases}
\end{equation}
where $N_{s}$ is the total number of neutrino sources, $F$ is the flux of each source, the slopes ($\alpha, \beta$) are fixed to (2.5, 1.5), $F_{\star}$ is the characteristic flux at the break, and $N_\star$ is the characteristic number of sources at $F_\star$ (see Ref.~\cite{Ando:2017xcb} for further details). The slope at high-flux regime ($\alpha = 2.5$) is generically true for any sources distributed homogeneously in the local volume where expansion of the Universe can be neglected. We fix the lower flux limit at $F_{0} = 0.1 F_{\star}$. It is justified since we are interested in the angular clustering of neutrino events, which is dominated only by sources with large fluxes. 

The mean of the flux distribution is related to the measured neutrino intensity $\Phi_\nu$, Eq.~(\ref{eq:spectra}), and can be found through
\begin{equation}
\int dF F \frac{dN_s}{dF} = 4\pi \Phi_\nu.
\end{equation}
We can analytically calculate the left-hand side using Eq.~(\ref{eq:1}) in order to obtain the characteristic flux, $F_{\star}$, as
\begin{equation}
F_{\star} = \frac{4 \pi \Phi_{\nu}}{N_{\star} \eta}\,,
\end{equation}
where $\eta = 2 + (1-F_{0}^{2-\beta})/(2-\beta)$. The source flux distribution is thus shaped by dividing the measured neutrino intensity over $N_{\star}$, and we leave $N_{\star}$ as a free parameter for the rest of the paper.

The all-sky map of the neutrino flux is obtained by randomly distributing the total number of sources, found by integrating Eq.~(\ref{eq:1}) over all fluxes, and assigning each source a random flux following the source flux distribution for fixed $N_{\star}$. 
Additionally, we consider the case of an infinite number of sources, the isotropic case, where we obtain the flux sky map by equally dividing the measured intensity over all pixels on the sky.

\subsection{Neutrino sky map simulation}
The number of neutrino detections per given energy range from single sources with flux $F$ is calculated by multiplying the flux by the exposure (the effective area multiplied by the live time) of the neutrino detector and integrating over the neutrino energy:
\begin{equation}
N_{\nu}= \int_{E_{\mu,{\rm th}}}^\infty dE_\mu \int_0^\infty dE_\nu \frac{dF}{dE_\nu} \frac{d\mathcal{E}(E_{\nu}, E_\mu, \theta)}{dE_\mu},
\end{equation}
where $d\mathcal{E}(E_{\nu}, E_{\mu},\theta) / dE_\mu$ is the differential exposure to yield muon energy proxy $E_\mu$ from the incident neutrino with energy $E_\nu$, which also depends on the zenith angle ($\theta$) due to the Earth absorption. The exposure data from IceCube is obtained from the previously mentioned URL. In the case of IceCube, we set a threshold on the muon energy proxy of $E_\mu > E_{\mu,{\rm th}} = 5\times 10^{4}$~GeV, while in the case of the KM3NeT exposure we set a threshold on the muon neutrino energy of $E_{\nu} > 100$~TeV. The KM3NeT effective area used in this analysis is at trigger level, and no cuts were applied in order to reduce background. Before analyzing the neutrino event map, we mask all pixels with $\cos\theta < 0.1$, where zenith angle $\theta$ represents the angle between the neutrino detectors' normal vector and the incoming neutrino direction.

\begin{figure}[!ht] 
\centering
\minipage{0.7\textwidth}
\includegraphics[width=\linewidth]{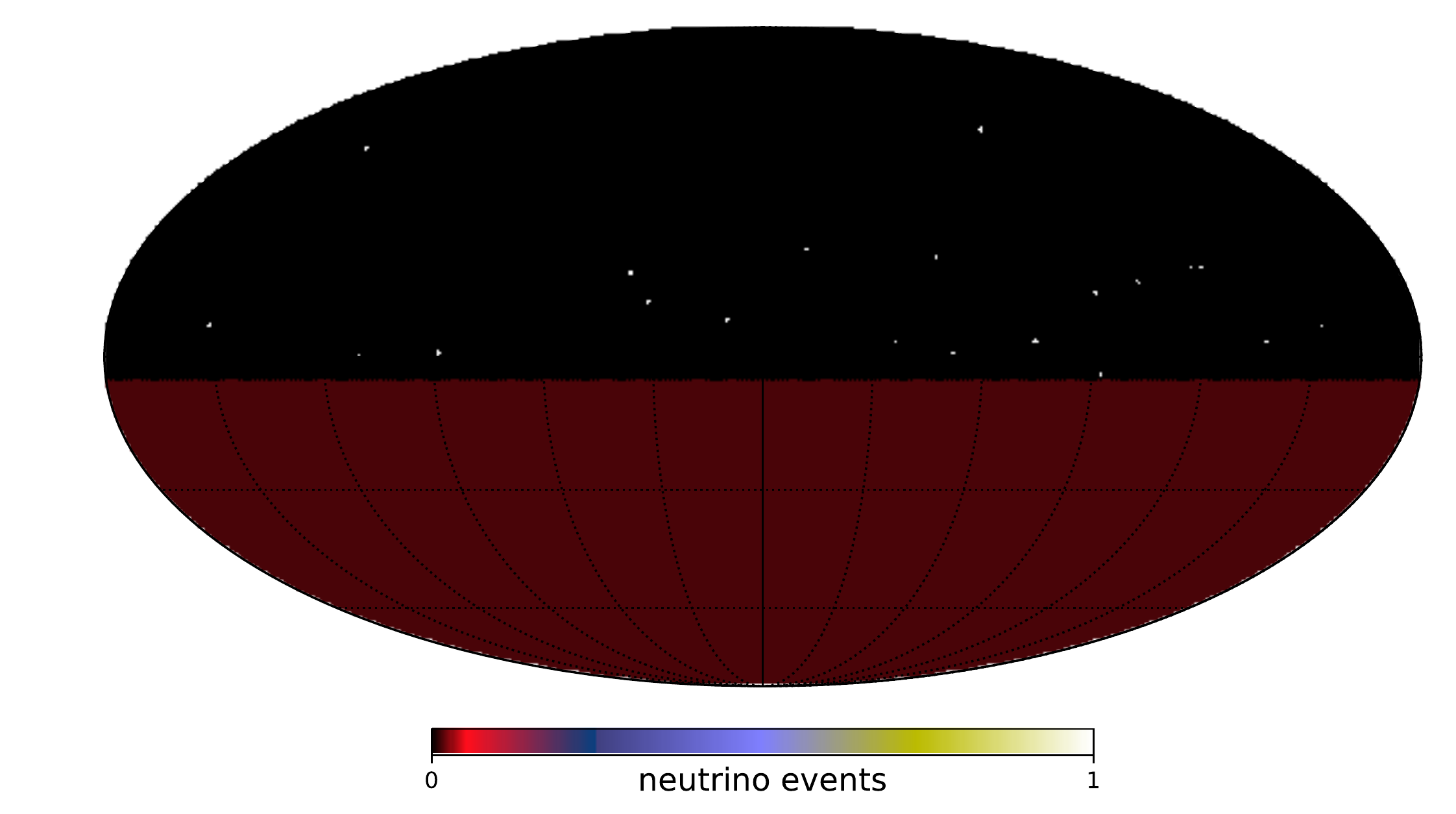}
\endminipage\hfill
\minipage{0.7\textwidth}
\includegraphics[width=\linewidth]{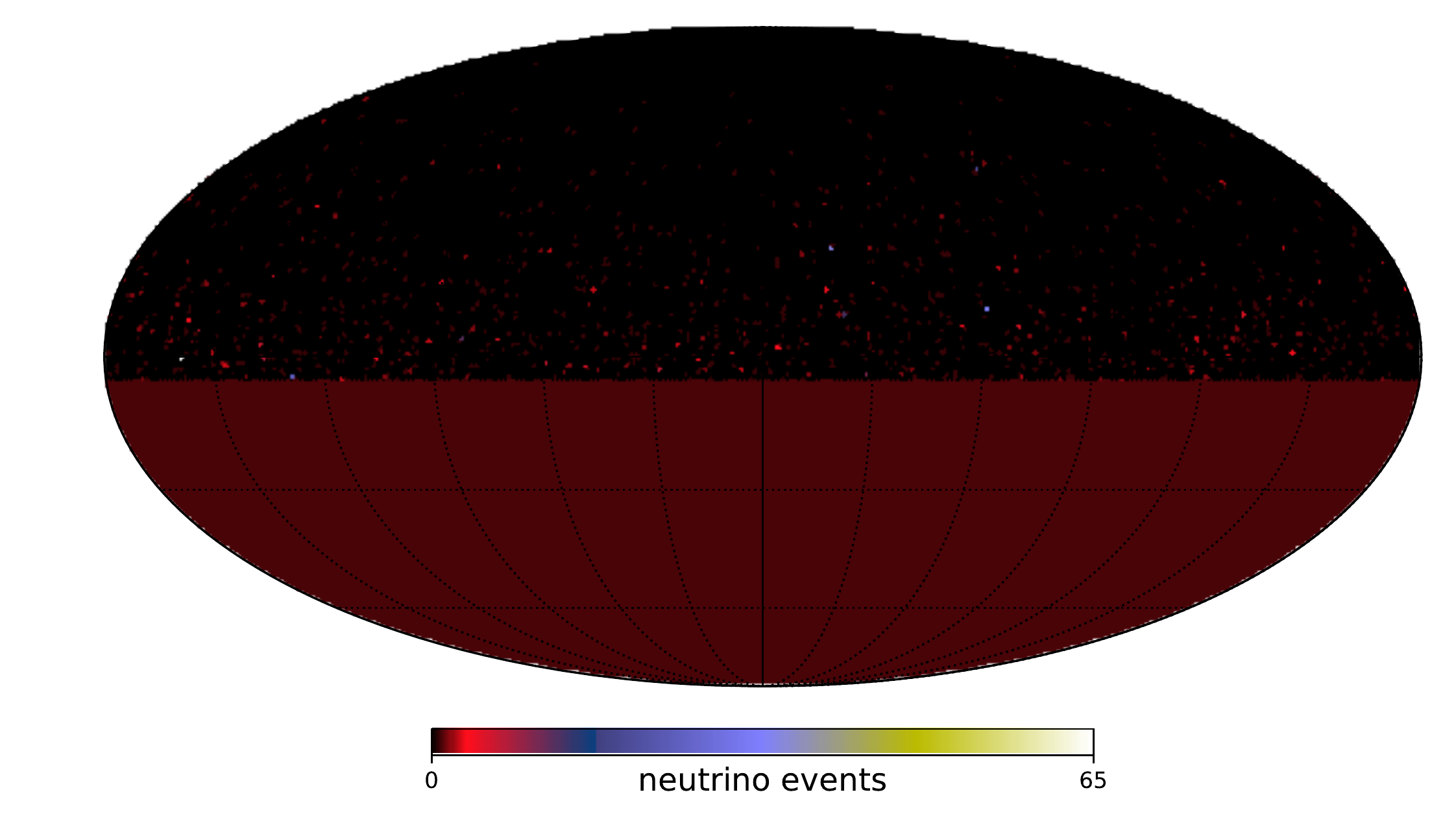}
\endminipage\hfill
\minipage{0.7\textwidth}%
\includegraphics[width=\linewidth]{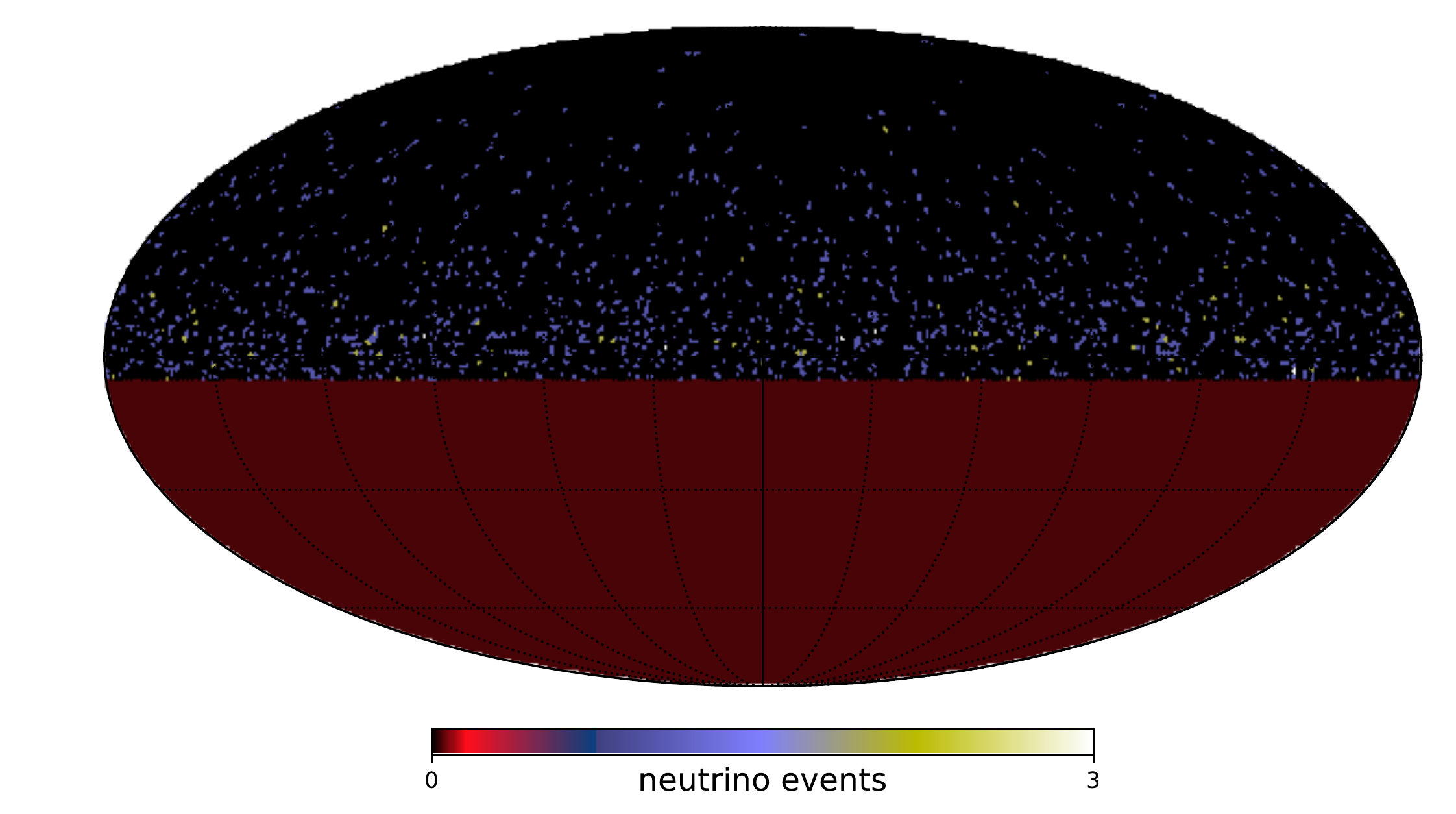}
\endminipage
\caption{Neutrino sky maps using IceCube exposure showing, neutrino events and the masked area in dark red. The top panel shows the 21 neutrino events above $E_{\mu}>50$ TeV from 2 years of IceCube observations. The middle and bottom panels show two simulated sky maps with $N_{\star} = 10^{3}$ and $N_{s} = \infty$, respectively, where the exposure is enhanced to 200 years for better illustration. }
\label{fig:skymap}
\end{figure}

Figure~\ref{fig:skymap} shows three examples of neutrino sky maps by using IceCube exposure, whereby the masked region is shown in dark red. The top panel shows the 21 neutrino events from 2 years of IceCube observations, which are isotropically distributed. The middle and bottom panels illustrate the angular clustering behavior by changing $N_{\star}$, showing $N_{\star} = 10^{3}$ and $N_{s} = \infty$, respectively, where the exposure is enhanced to 200 years for better interpretation. The sky map with $N_{\star} = 10^{3}$ has brighter sources, which can be seen by the maximum number of events coming from one pixel being 65. While having around the same total number of neutrino events, the sky map for $N_{s} = \infty$ shows a maximum of 3 events per pixel due to the sources being very faint. The measured distribution of the events can thus be used to constrain the contribution of bright sources to the intensity due to their large fluctuations, and we will assess the parameter $N_{\star}$ in this analysis. 

\section{Method}\label{sec:2}
\subsection{Auto-correlation angular power spectrum}
The anisotropy of the neutrino events can be assessed through the angular power spectrum (APS), which represents the fluctuations as a function of the angular scale. Hereby, the neutrino event sky map is expanded into spherical harmonics as follows:
\begin{equation}
N\left(\theta, \phi \right) = \sum_{\ell m} a_{\ell m} Y_{\ell m}\left(\theta, \phi \right),
\end{equation}
where $N\left(\theta, \phi \right)$ is the neutrino events at their latitude ($\theta$) and longitude ($\phi$), and $Y_{\ell m}(\theta, \phi)$ are the spherical harmonic functions. The APS is described by averaging the expansion coefficients over the sky:
\begin{equation}
C_{\ell} = \frac{1}{2\ell +1} \sum_{m=-\ell}^{\ell} |a_{\ell m}|^2 .
\end{equation}
We computed the APS using \texttt{anafast} tool from the software package HEALPix~\cite{HealPix}. 

\subsection{Analysis}
We perform the simulations for 50 logarithmically-spaced values of $N_{\star}$, in the range between 10 and $10^{6}$, and additionally for the isotropic case of $N_{s}=\infty$. A set of $10^{5}$ (IceCube, IceCube-Gen2) and $10^{4}$ (KM3NeT) Monte Carlo simulations are performed for each parameterization of the sky map. 
We adopt the angular resolution for the neutrino detectors as $\sigma=0.5^{\circ}$ for IceCube, $\sigma=0.3^{\circ}$ for IceCube-Gen2 ~\cite{IceCubeGen2}, and $\sigma=0.07^{\circ}$ for KM3NeT. These resolutions allow a maximum multipole moment of $\ell_{max} = 192$, 329, and 1400, respectively. The specifications are summarized in Table~\ref{Tab:para}.

\begin{figure}[h!]
	\centering
	\includegraphics[width=0.55\textwidth]{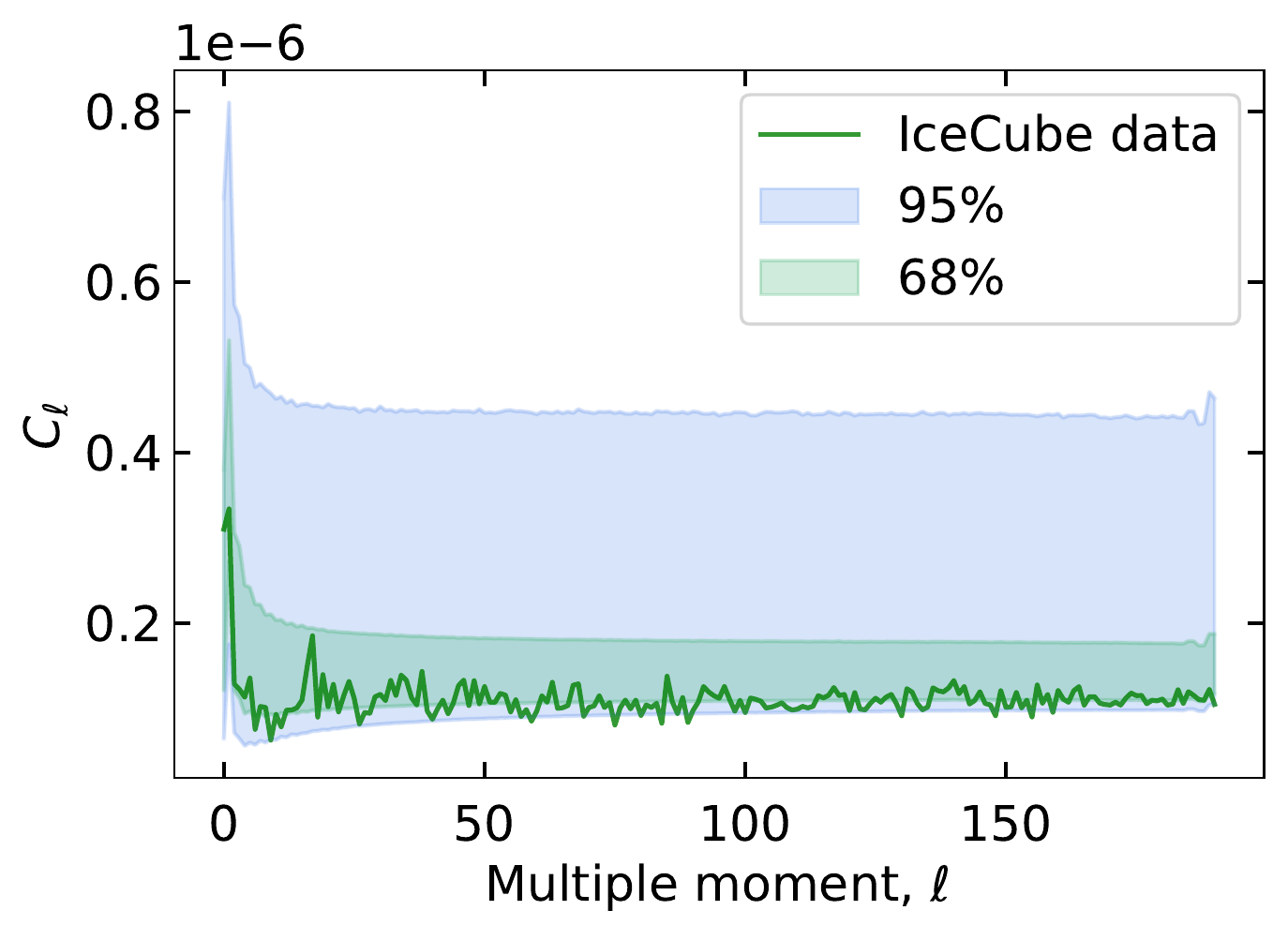}
	\caption{The angular power spectrum with two years of IceCube exposure, showing the $95\%$ and $68\%$ containment bands of the $10^{5}$ simulated sky maps for $N_{\star}=10^{2}$, as well as the APS of the IceCube observation (solid line).}
	\label{fig:cl_example} 
\end{figure}

The APS of $10^{5}$ MC simulations is illustrated in Fig.~\ref{fig:cl_example} in the case of $N_{\star}=10^{2}$ with two years of IceCube exposure, showing the $95\%$ and $68\%$ containment bands as well as the APS of the observed sky map (solid line). The figure clearly shows that, in the case of $N_\star = 10^2$, the probability distribution of $C_\ell$ cannot be well represented by the Gaussian, being highly skewed towards higher values of $C_\ell$.
In order to see this more systematically, and more importantly to discuss detectability of sources using the APS, we adopt the following two analysis methods.

\begin{table}[!ht]
	\begin{center}
	\caption{The IceCube, IceCube-Gen2 and KM3NeT specifications applied for the simulation.}
    \label{Tab:para}
		\begin{tabular}{ | c | c | c | c | c | }
        \hline
			 & Exposure time [yr] & Angular resolution [deg] & $\ell_{max}$ & Nr. of simulations \\ \hline
			IceCube      &   2 & 0.5 & 192 & $10^{5}$ \\ 
			IceCube-Gen2 &	10 & 0.3 & 329 & $10^{5}$ \\ 
			KM3NeT       &	10 & 0.07& 1400 & $10^{4}$ \\ 
            \hline
		\end{tabular}
	\end{center}
\end{table}

\paragraph{$\bar{C_{\ell}}$ distribution.}
The first method that we approach is by distributing the mean APS for various values of $N_{\star}$ and compare their differences in shape and normalizations. We therefore take the unweighted average of $C_\ell$, denoted as $\bar{C_{\ell}}$, over the multipole range $50 \le \ell \le \ell_{max}$, where the first 50 multipole moments are removed to reduce uncertainties at large angular scales.
We discuss $\bar C_\ell$ for $N_{\star} = 10$, $10^{2}$, $10^{3}$, $10^{4}$ and $10^{5}$. This method assumes no prior neutrino sky map information and therefore allows for a direct comparison with the APS of the observed neutrino sky. 

\paragraph{$p$-value.}
With the second approach, we fit the APS of the observed sky map to the APS of the simulated sky maps in order to test if $N_{\star}$ is compatible, and we use all multipole information. We consider the following $\chi^{2}$, obtained for each $N_{\star}$:
\begin{equation}
\chi^{2} \left(C_{\ell}\right) = \sum_{\ell \ell'} \left(C_{\ell} - C_{\ell}^{\rm{mean}}\right) (\rm{Cov}_{\ell\ell'})^{-1} \left(C_{\ell'} - C_{\ell'}^{\rm{mean}}\right),
\label{eq:chi2}
\end{equation}
where $C_{\ell}$ is the simulated APS, $C_{\ell}^{\rm{mean}}$ is the mean value at each $\ell$, and $\rm{Cov}_{\ell\ell'}$ is the covariance matrix, where the latter two are obtained from the set of simulations.
For each $N_\star$, we calculate the probability density function (PDF) of $\chi^2$, $P(\chi^2|N_\star)$, as well as $C_\ell^{\rm mean}$ and ${\rm Cov}_{\ell\ell'}$. 
We then compare the value of $\chi_{\rm data}^2\equiv \chi^2(C_{\ell}^{\rm{data}})$ obtained from the APS of the observed sky map, $C_\ell^{\rm data}$, and quantify the probability of obtaining the same or more extreme values (towards either greater or smaller direction) of $\chi^{2}$ for each $N_{\star}$, noted as the $p$-value, as follows:
\begin{equation}
p = \min\left[\int_{\chi^2_{\rm data}}^\infty d\chi^2 P(\chi^2|N_\star), \int_0^{\chi^2_{\rm data}} d\chi^2 P(\chi^2|N_\star)\right].
\end{equation}
In the following, we look for constraints at 95\% confidence level (CL), which is equivalent to $p = 0.05$.

\section{Results}\label{sec:3}
Following the methodology summarized in the previous section, we analyze the APS of $10^{5}$ and $10^{4}$ simulations for each parameterization of $N_{\star}$ using IceCube, IceCube-Gen2 and KM3NeT specifications.

\subsection{\texorpdfstring{$\bar{C_{\ell}}$}{}  distributions}
\subsubsection{IceCube}
\begin{figure}[!ht] 
\centering
\minipage{0.5\textwidth}
\includegraphics[width=\linewidth]{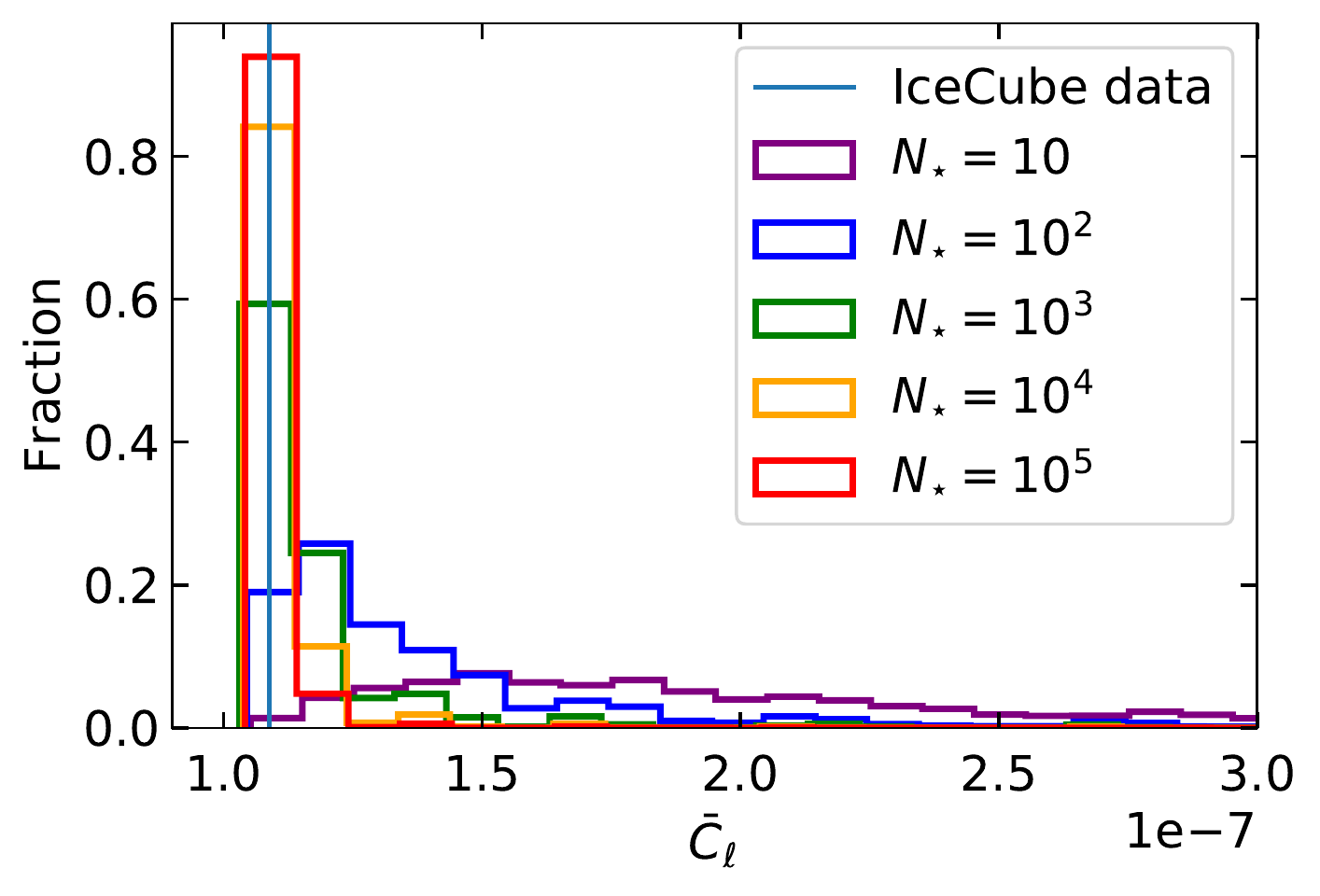}
\endminipage\hfill
\minipage{0.5\textwidth}
\includegraphics[width=\linewidth]{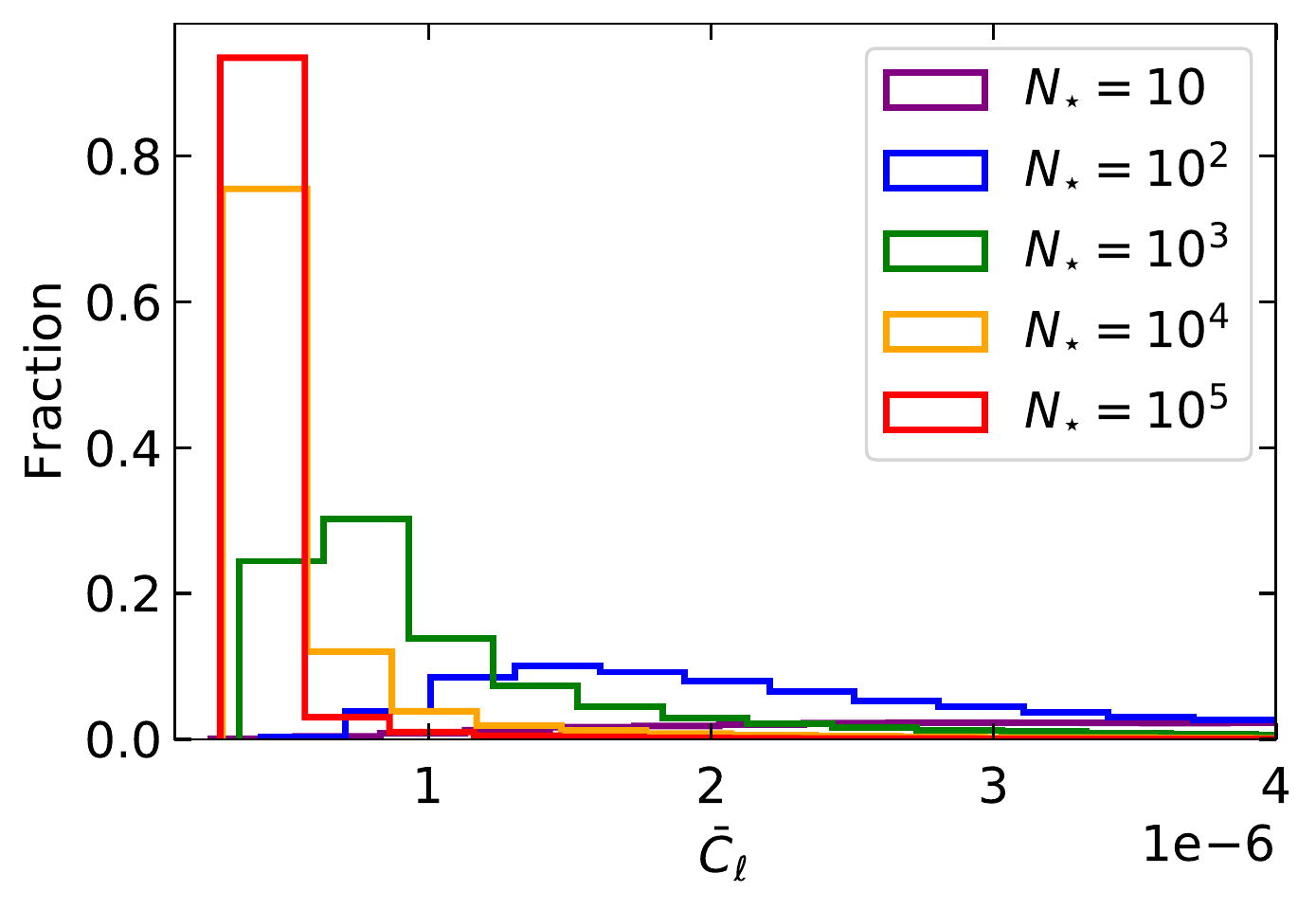}
\endminipage\hfill
\minipage{0.5\textwidth}%
\includegraphics[width=\linewidth]{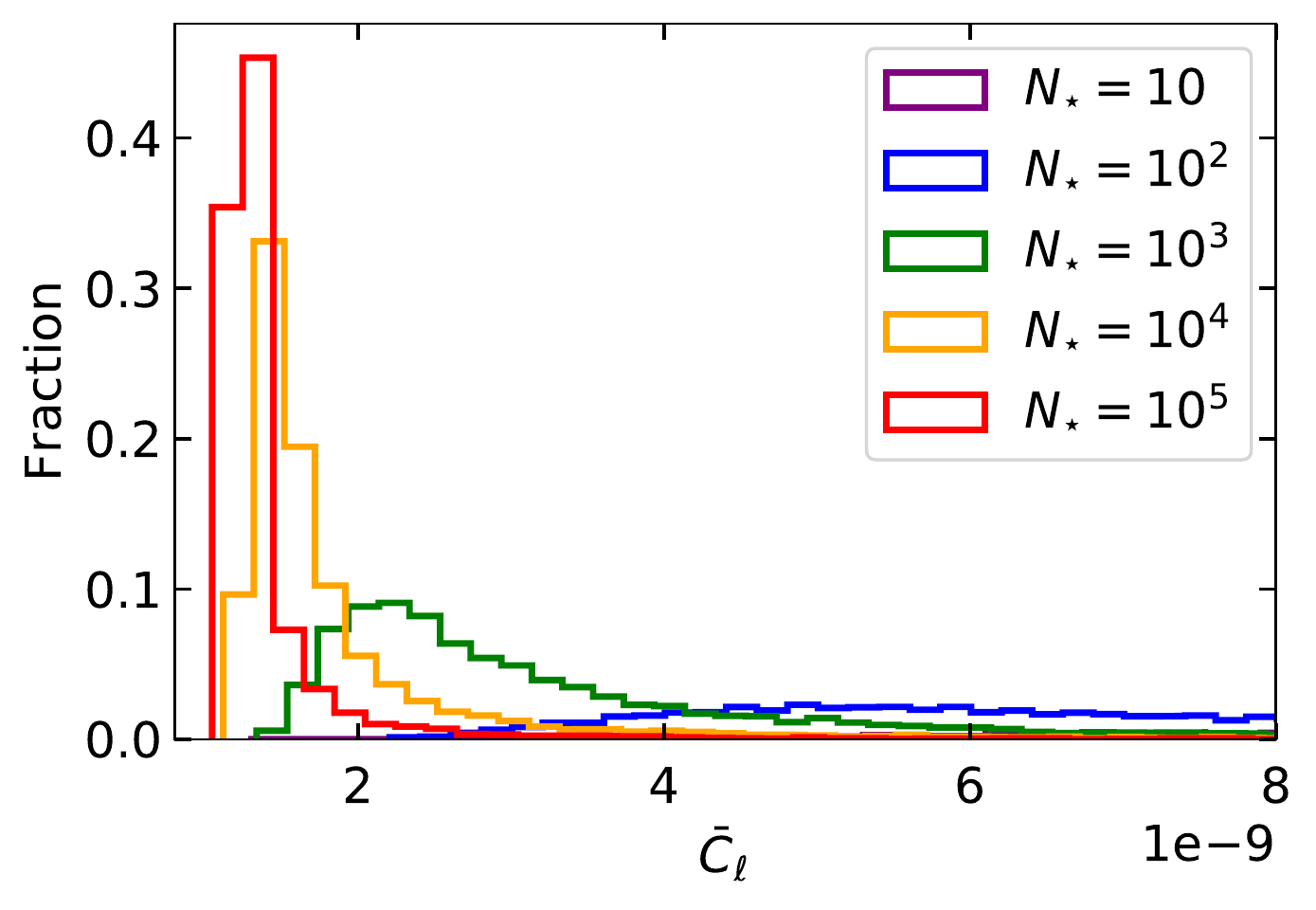}
\endminipage
\caption{$\bar{C_{\ell}}$ distributions with $N_{\star} = 10, 10^{2}, 10^{3}, 10^{4}$ and $10^{5}$. The top left panel shows the distributions using two years of IceCube exposure, as well as $\bar{C_{\ell}}$ obtained from the observed sky map, illustrated as the vertical line. The distributions in the top right panel are obtained with ten years of IceCube-Gen2 exposure and the distributions in the bottom bottom panel with ten years of KM3NeT exposure. 
The values of $\bar{C_\ell}$ using KM3NeT exposure are much smaller than those for IceCube and IceCube-Gen2 because of much finer pixel size.}
\label{fig:distr}
\end{figure}

The top left panel of Fig.~\ref{fig:distr} shows the result of the $\bar{C_{\ell}}$ distributions for $N_{\star} = 10,$ $10^{2}$, $10^{3}$, $10^{4}$ and $10^{5}$ using two years of IceCube exposure. The distributions for smaller $N_{\star}$ deviate from Gaussian showing a tail towards larger $\bar C_\ell$. 
The $\bar{C_{\ell}}$ obtained from the observed sky map is also illustrated, as the vertical blue line, which has the value $\bar{C_{\ell}}^{\rm data} = 1.09\times 10^{-7}$~sr. Comparing with the exclusion values in Table~\ref{Tab:1}, we can exclude $N_{\star} = 10$ with 95\%CL. 

\begin{table}[!ht]
	\begin{center}
	\caption{Upper and lower limits on $\bar{C}_{\ell}$ at 90\%, 95\% and 99\% CL using two years of IceCube exposure. When comparing with the observed value $\bar{C}_{\ell}^{\rm{data}}=1.09\cdot10^{-7} $, we can exclude $N_{\star}=10$ with 95\% CL.}
    \label{Tab:1}
		\begin{tabular}{ | c | c | c | c | }
        \hline
			$N_{\star}$ &	90\% CL & 95\% CL & 99\% CL \\ \hline
			$10$     &	$1.20\cdot 10^{-7}$--$6.80\cdot 10^{-7}$ & $1.19\cdot 10^{-7}$--$9.22\cdot 10^{-7}$ & $1.09\cdot 10^{-7}$--$1.59\cdot 10^{-6}$\\ 
			$10^{2}$ &	$1.09\cdot 10^{-7}$--$2.97\cdot 10^{-7}$ & $1.08\cdot 10^{-7}$--$4.33\cdot 10^{-7}$ & $1.07\cdot 10^{-7}$--$1.06\cdot 10^{-6}$\\ 
			$10^{3}$ &	$1.08\cdot 10^{-7}$--$1.71\cdot 10^{-7}$ & $1.08\cdot 10^{-7}$--$2.14\cdot 10^{-7}$ & $1.07\cdot 10^{-7}$--$5.71\cdot 10^{-7}$\\ 
			$10^{4}$ &	$1.08\cdot 10^{-7}$--$1.21\cdot 10^{-7}$ & $1.07\cdot 10^{-7}$--$1.41\cdot 10^{-7}$ & $1.07\cdot 10^{-7}$--$2.66\cdot 10^{-7}$\\ 
			$10^{5}$ &	$1.08\cdot 10^{-7}$--$1.19\cdot 10^{-7}$ & $1.07\cdot 10^{-7}$--$1.20\cdot 10^{-7}$ & $1.07\cdot 10^{-7}$--$1.71\cdot 10^{-7}$\\ 
            \hline
		\end{tabular}
	\end{center}
\end{table}

\begin{table}[!ht]
	\begin{center}
	\caption{Upper and lower limits on $\bar{C}_{\ell}$ at 90\%, 95\% and 99\% CL using ten years of IceCube-Gen2 exposure.}
    
    \label{Tab:2}
		\begin{tabular}{ | c | c | c | c | }
        \hline
			$N_{\star}$ &	90\% CL & 95\% CL & 99\% CL \\ \hline
			$10$     &	$1.84\cdot 10^{-6}$--$1.60\cdot 10^{-4}$ & $1.41\cdot 10^{-6}$--$3.90\cdot 10^{-4}$ & $8.27\cdot 10^{-7}$--$3.43\cdot 10^{-3}$\\ 
			$10^{2}$ &	$1.05\cdot 10^{-6}$--$3.47\cdot 10^{-5}$ & $9.22\cdot 10^{-7}$--$8.13\cdot 10^{-5}$ & $7.41\cdot 10^{-7}$--$6.93\cdot 10^{-4}$\\ 
			$10^{3}$ &	$4.84\cdot 10^{-7}$--$7.93\cdot 10^{-6}$ & $4.54\cdot 10^{-7}$--$1.89\cdot 10^{-5}$ & $4.06\cdot 10^{-7}$--$1.52\cdot 10^{-4}$\\ 
			$10^{4}$ &	$3.30\cdot 10^{-7}$--$1.93\cdot 10^{-6}$ & $3.21\cdot 10^{-7}$--$4.02\cdot 10^{-6}$ & $3.07\cdot 10^{-7}$--$3.20\cdot 10^{-5}$\\ 
			$10^{5}$ &	$2.95\cdot 10^{-7}$--$6.58\cdot 10^{-7}$ & $2.91\cdot 10^{-7}$--$1.14\cdot 10^{-6}$ & $2.82\cdot 10^{-7}$--$7.06\cdot 10^{-6}$\\ 
            \hline
		\end{tabular}
	\end{center}
\end{table}

\begin{table}[!ht]
	\begin{center}
	\caption{Upper and lower limits on $\bar{C}_{\ell}$ at 90\%, 95\% and 99\% CL using ten years of KM3NeT exposure.}
    \label{Tab:3}
		\begin{tabular}{ |c | c | c | c| }
        \hline
			$N_{\star}$ &	90\% CL & 95\% CL & 99\% CL \\ \hline
			$10$     &	$7.45\cdot 10^{-9}$--$5.37\cdot 10^{-7}$ & $5.93\cdot 10^{-9}$--$1.33\cdot 10^{-6}$ & $3.52\cdot 10^{-9}$--$1.06\cdot 10^{-5}$ \\ 
			$10^{2}$ &	$3.69\cdot 10^{-9}$--$1.05\cdot 10^{-7}$ & $3.24\cdot 10^{-9}$--$2.40\cdot 10^{-7}$ & $2.73\cdot 10^{-9}$--$1.47\cdot 10^{-6}$ \\ 
			$10^{3}$ &	$1.76\cdot 10^{-9}$--$2.56\cdot 10^{-8}$ & $1.67\cdot 10^{-9}$--$5.81\cdot 10^{-8}$ & $1.52\cdot 10^{-9}$--$4.75\cdot 10^{-7}$ \\ 
			$10^{4}$ &	$1.28\cdot 10^{-9}$--$6.15\cdot 10^{-9}$ & $1.25\cdot 10^{-9}$--$1.26\cdot 10^{-8}$ & $1.19\cdot 10^{-9}$--$9.76\cdot 10^{-8}$ \\ 
			$10^{5}$ &	$1.16\cdot 10^{-9}$--$2.43\cdot 10^{-9}$ & $1.15\cdot 10^{-9}$--$4.04\cdot 10^{-9}$ & $1.12\cdot 10^{-9}$--$1.96\cdot 10^{-8}$ \\ 
            \hline
		\end{tabular}
	\end{center}
\end{table}

\subsubsection{IceCube-Gen2}
Fluctuations caused by bright sources are expected to become dominant if we increase the exposure, and we expect the distributions to start showing different shapes. Indeed, in the case for ten years of IceCube-Gen2 exposure, as shown in the top right panel of Fig.~\ref{fig:distr}, we find that the distributions are significantly separated from each other, allowing to constrain $N_{\star}$ when comparing with future observations. Upper and lower limits on $\bar{C_{\ell}}$ are obtained with 90\%, 95\% and 99\% CL, and are represented in Table~\ref{Tab:2}. If the measured $\bar C_\ell$ are beyond the ranges shown in the table, one can exclude the model with $N_\star$ at a given CL.

\subsubsection{KM3NeT}
The bottom panel of Fig,~\ref{fig:distr} shows the $\bar{C_{\ell}}$ distributions using ten years of KM3NeT exposure, where we find that the distributions are again significantly distinguishable from each other in order to constrain $N_{\star}$ when comparing with future $\bar{C_{\ell}}^{\rm data}$. In table~\ref{Tab:3}, we present the corresponding exclusion limits on $\bar{C_{\ell}}$. The values of $\bar{C_{\ell}}$ are much smaller than those for IceCube and IceCube-Gen2 due to the higher detector resolution, and thus finer pixel size.

\subsection{Correlation between multipoles}

\begin{figure}[!ht]
	\centering
	\begin{minipage}{0.49\textwidth}
		\centering
		\includegraphics[width=.9\linewidth]{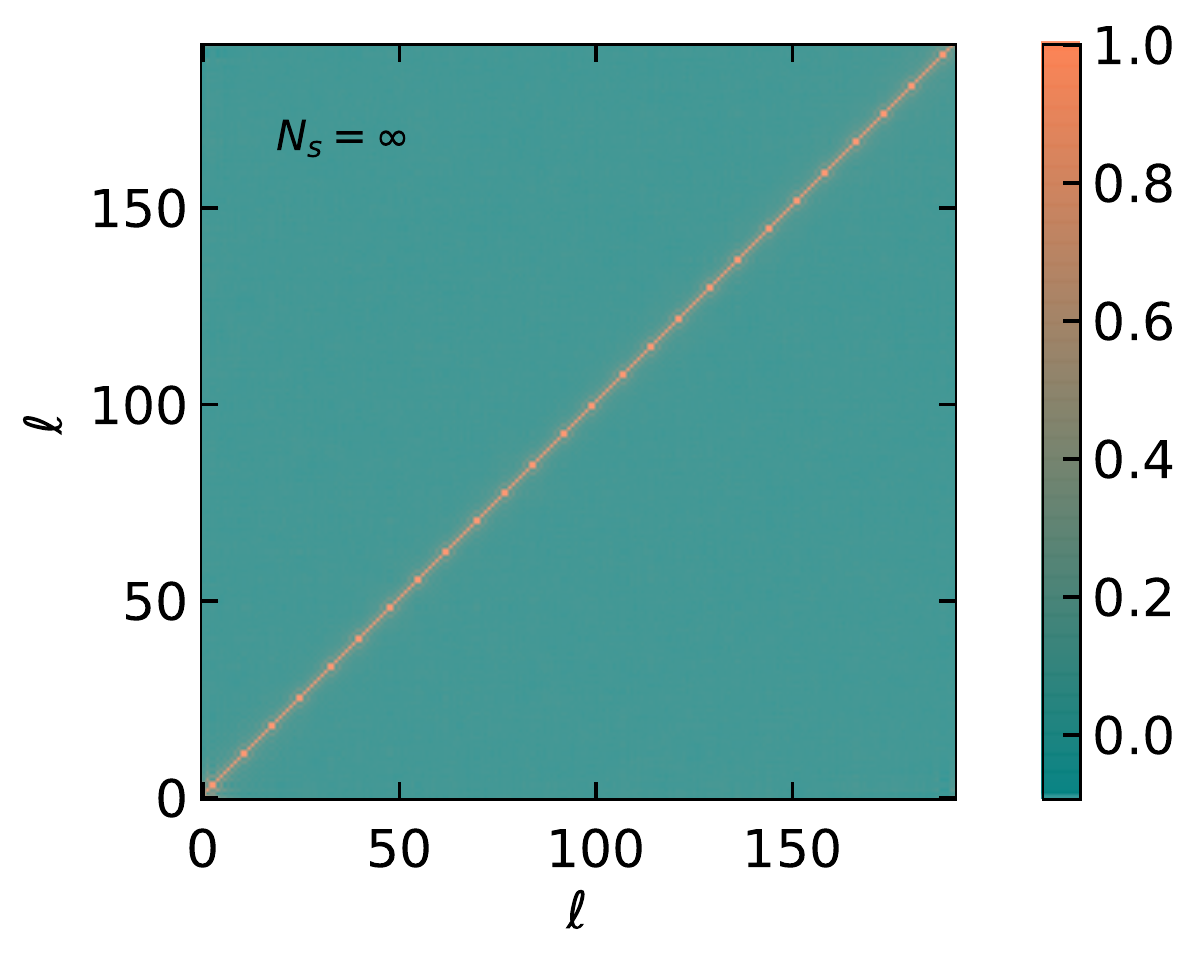}
	\end{minipage}
	\begin{minipage}{0.49\textwidth}
		\centering
		\includegraphics[width=.9\linewidth]{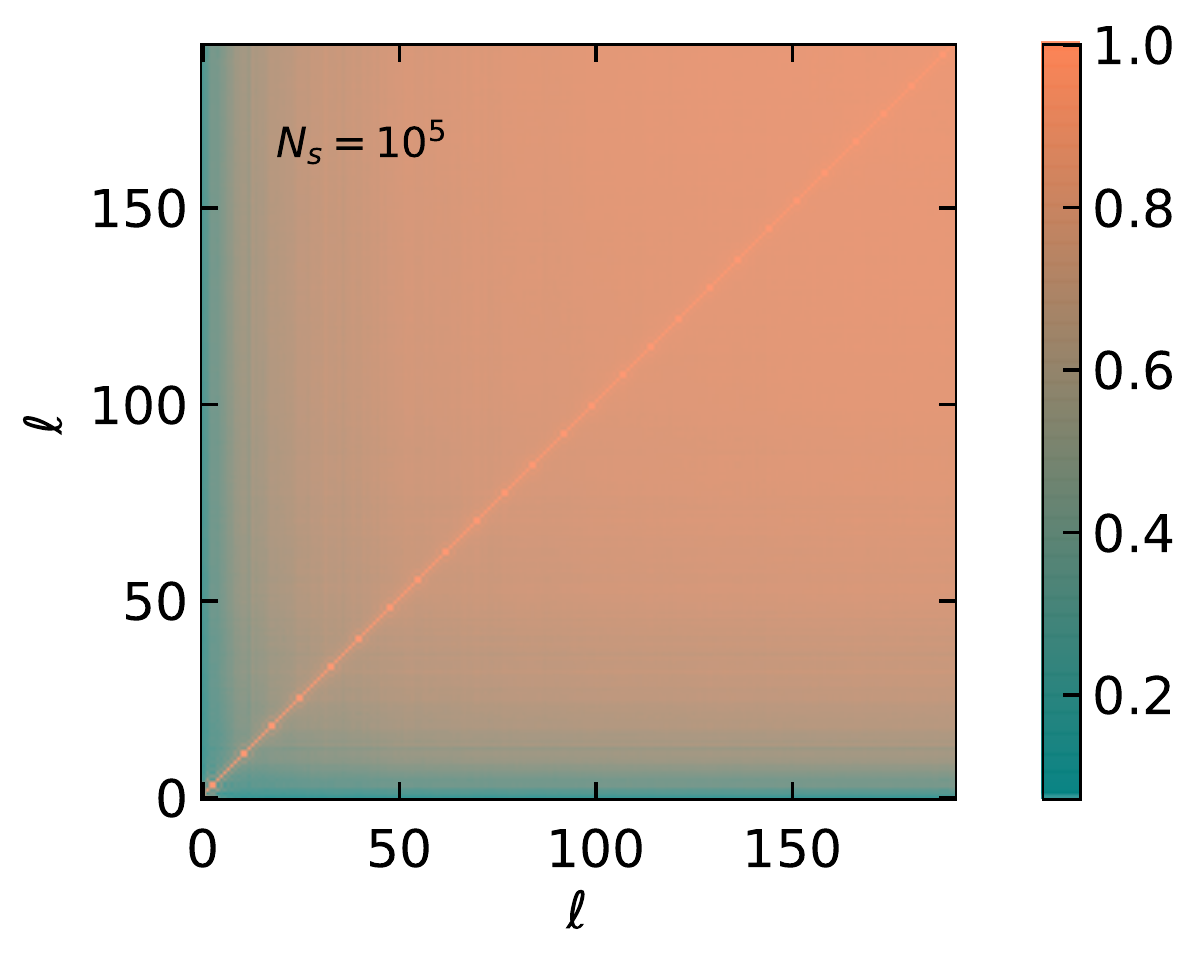} 
	\end{minipage}
	\begin{minipage}{0.49\textwidth}
		\centering
		\includegraphics[width=.9\linewidth]{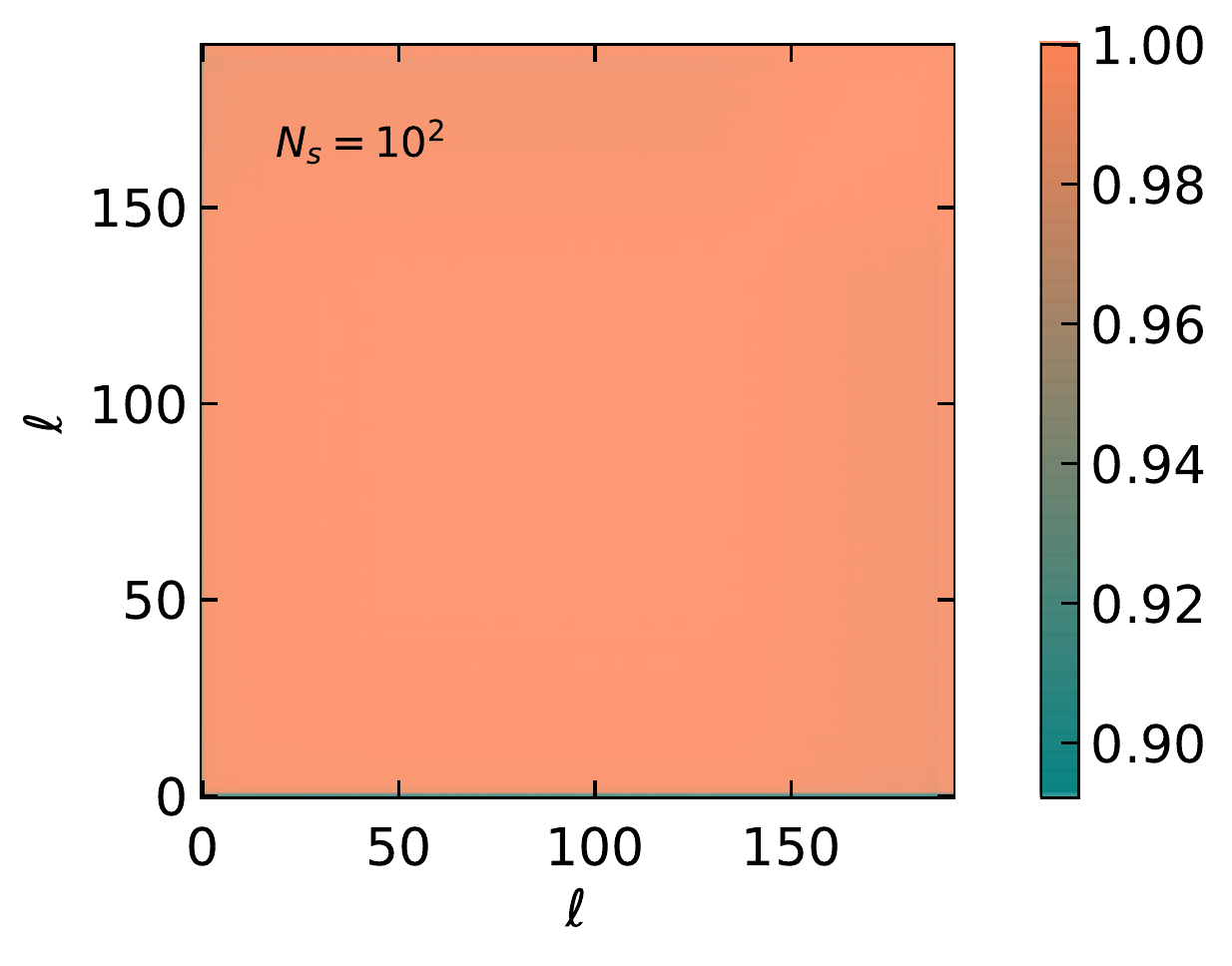} 
	\end{minipage}
	\caption{Correlation matrices using two years of IceCube exposure in the cases of $N_{s} = \infty$ (left top), $N_{\star} = 10^{5}$ (right top) and $N_{\star} = 10^{2}$ (bottom).}
\label{fig:cor}
\end{figure}

In order to estimate the $p$-value from the $\chi^2$ analysis (see, Eq.~\ref{eq:chi2}), we need to evaluate the covariance matrix ${\rm Cov}_{\ell\ell'}$.
The covariance matrix is a $\ell _{\rm max} \times \ell _{\rm max}$ matrix where the diagonal part indicates the standard deviation at each multipole, while the off-diagonal elements represent the correlation between different multipole moments.
Especially if the events are dominated by small number of sources, $N_\star$, it is likely that the different multipoles are highly correlated with each other, showing non-negligible off-diagonal components (e.g., \cite{Ando:2017wff}).
We obtain the covariance matrix 
\begin{equation}
\rho_{\ell\ell'} \equiv \frac{{\rm Cov}_{\ell\ell'}}{\sqrt{{\rm Cov}_{\ell\ell}{\rm Cov}_{\ell'\ell'}}},
\end{equation}
of the APS for each $N_{\star}$ using the Monte Carlo simulations.  We find that the correlation depends on $\ell$ and $N_{\star}$, as illustrated in Fig.~\ref{fig:cor}, which shows the correlation matrix $\rho_{\ell\ell'}$ using two years of IceCube exposure in the cases of $N_{s} = \infty$ (top left), $N_{\star} = 10^{5}$ (top right) and $N_{\star} = 10^{2}$ (bottom). We notice that the isotropic case does not show any correlation between different multipole moments, while decreasing number of sources show an increase in correlation. This implies that, especially at small angular scales, we see a deviation between different $N_{\star}$, and thus shows the importance of a small angular resolution using this analysis.  It also shows the importance of treating the full covariance matrix rather than the diagonal components, as doing the latter will introduce systematic bias of the obtained constraints especially for small values of $N_\star$, which highlight an important difference from the earlier work~\cite{Ando:2017xcb}.

\subsection{$p$-value analysis}

\subsubsection{IceCube}

\begin{figure}[h!]
	\centering
	\includegraphics[width=0.8\textwidth]{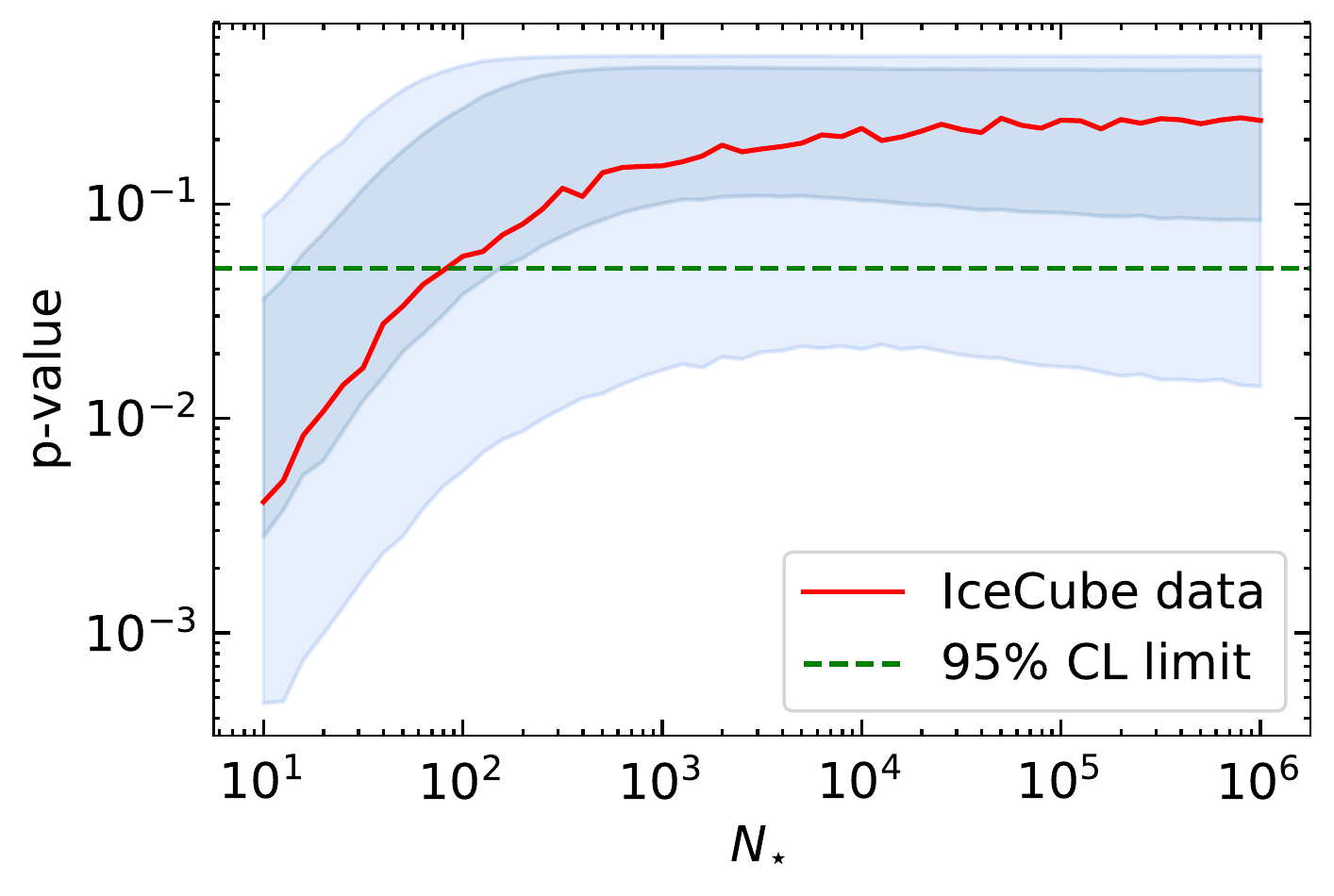}
	\caption{The $p$-value obtained by fitting the APS from the simulated sky maps using $N_{\star}$ to the APS from two years of IceCube data, shown as a red solid curve. Also illustrated are the expected $1\sigma$ and $2\sigma$ containment bands, where the simulated isotropic sky is used as mock data, shown as the shaded areas in blue and light blue respectively, and the horizontal green line represents the exclusion limit.}
	\label{fig:pval_2} 
\end{figure}

We tested the $N_{\star}$ parameterizations by fitting the APS of the simulated sky map to the APS of the observed sky map with 21 detections from two years of IceCube observations~\cite{Aartsen:2015rwa}, and the result is presented in Fig.~\ref{fig:pval_2} as a red solid curve. Moreover, the blue and light-blue shaded regions show the expected $1\sigma$ and $2\sigma$ containment bands, respectively, from the analysis using the APS of the simulated isotropic neutrino sky as mock data, and the green horizontal line shows the exclusion limit with $p = 0.05$. Following this exclusion line, we find a lower limit of $N_{\star} > 82$ at 95\% CL, which is well within the $1\sigma$ exclusion region using the simulated isotropic sky. This exclusion limit corresponds to very rare and bright sources, which are indeed not expected to dominate since the distribution of the 21 events are consistent with an isotropic expectation. 

If we still measure an isotropic sky with 10 years of IceCube exposure, we can exclude $N_{\star} = 10$--$4\times10^{3}$ with 95\% CL, where the region is the $1\sigma$ exclusion region.

\subsubsection{IceCube-Gen2}
IceCube-Gen2 is expected to have an effective area of a factor of ten times larger than IceCube, and we assessed its sensitivity using ten years of exposure. We first assumed that we keep on measuring an isotropic sky in the future, with mock data $N_{s} = \infty$, and find the $p$-values presented in the top left panel of Fig.~\ref{fig:pval_future}, where the shaded regions are the $1\sigma$ and $2\sigma$ containment bands from the $10^{5}$ analyzed APS, and the horizontal green line represents the exclusion limit of $p= 0.05$. We find an exclusion of bright sources below $N_{\star} = 10^{4}$--$2\times 10^{5}$, where the range represents the $1\sigma$ band.

If, however, we do measure clustering of events in the future due to bright sources, the contribution of less bright sources can be constrained and the exclusion tendencies will change. The middle and bottom left panels of Fig.~\ref{fig:pval_future} show the results by assuming $N_{\star} = 10^{4}$ and $N_{\star} = 10^{2}$, respectively, where the shaded areas and green line represent the aforementioned values. Indeed, in these cases we can significantly exclude the contribution of weak sources with large large $N_{\star}$, as well as very bright sources. Assuming that $N_{\star} = 10^{2}$, we can exclude source populations with $N_{\star}$ larger than $4\times10^{3}$--$9\times10^{3}$, as well as very rare, bright sources with $N_{\star}$ below $10$--$60$ with $95\%$ CL, where the range represents the $1\sigma$ band. Assuming $N_{\star}=10^{4}$, we can constrain rare, bright source populations even further, showing an $1\sigma$ exclusion band of $N_{\star} = 4\times10^{2}$--$6\times10^{3}$, while we also find an upper exclusion limit with $N_{\star}$-values larger than $6\times 10^{4}$.

\subsubsection{KM3NeT}
\begin{figure}[h!]
  \centering
  \includegraphics[width=.45\textwidth]{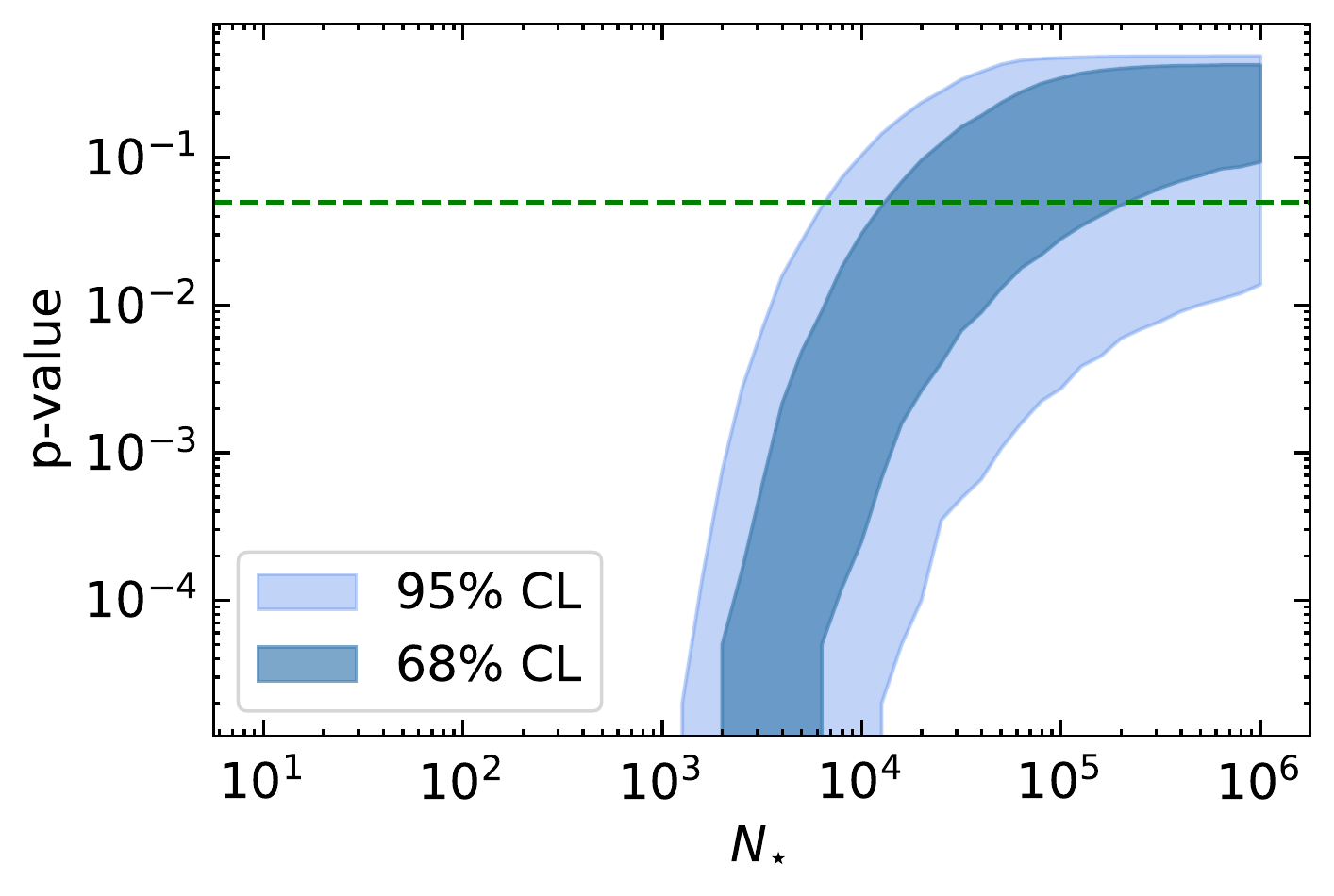}
  \hspace{1cm}
  \includegraphics[width=.45\textwidth]{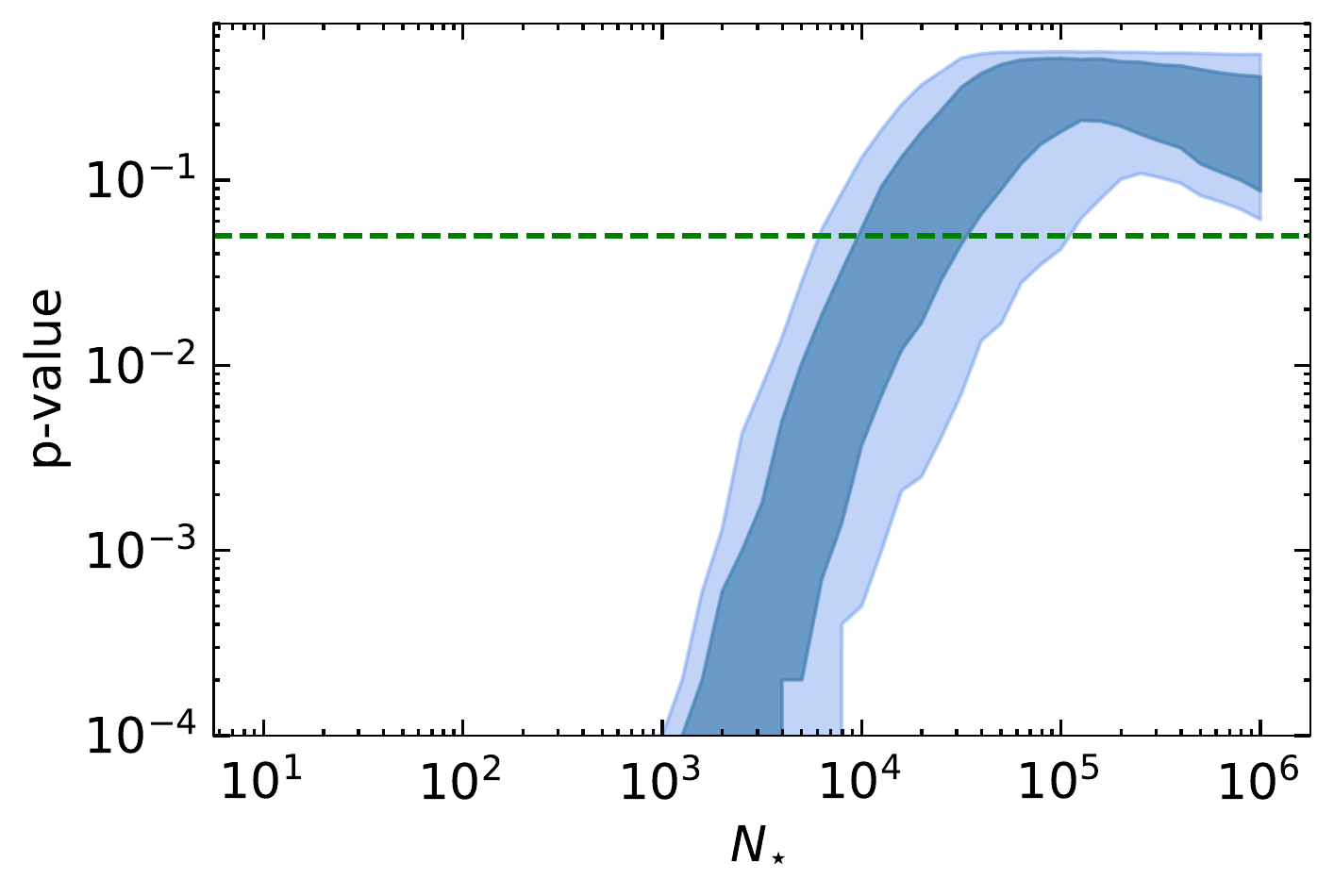}
  \includegraphics[width=.45\textwidth]{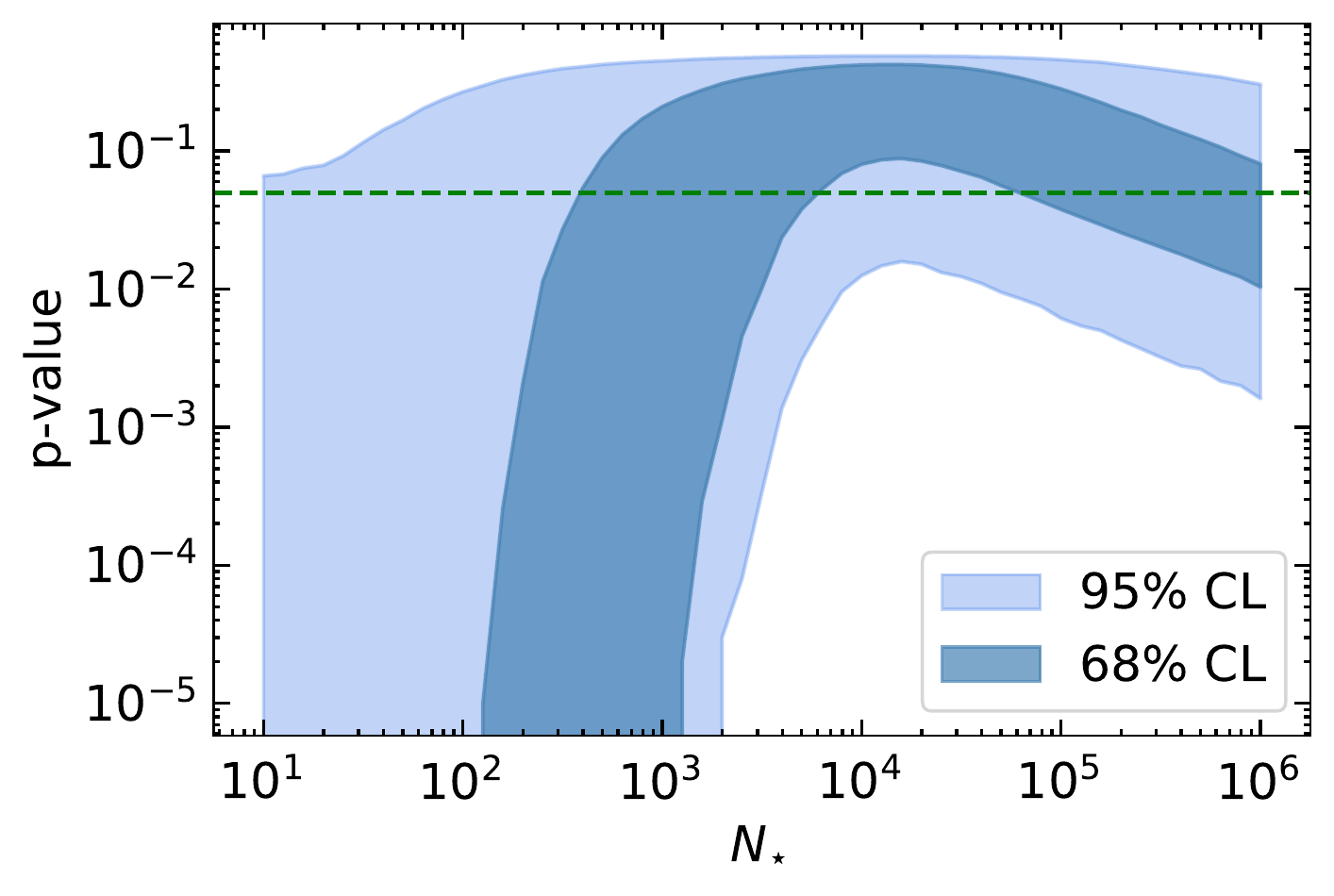}
  \hspace{1cm}
  \includegraphics[width=.45\textwidth]{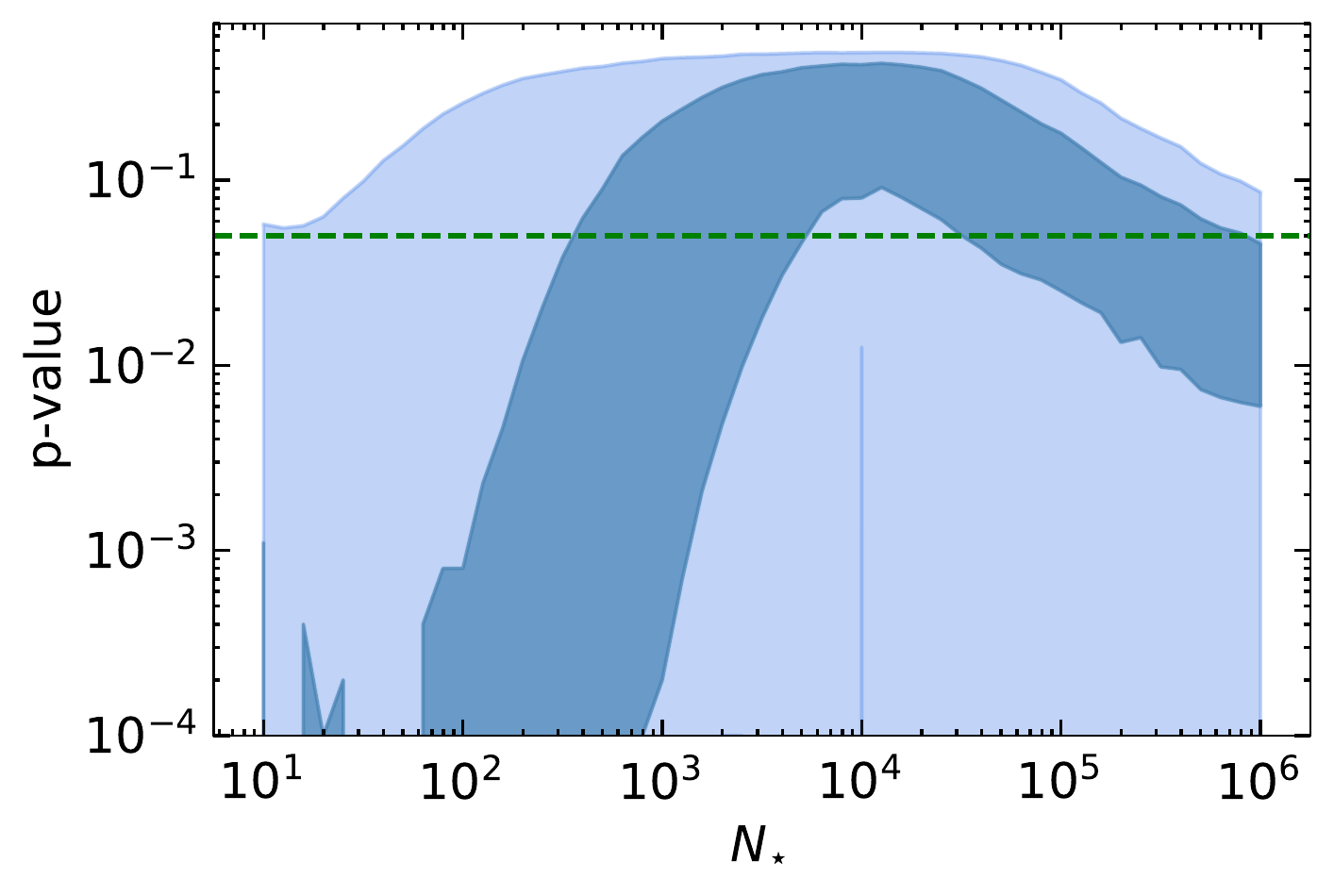}
  \includegraphics[width=.45\textwidth]{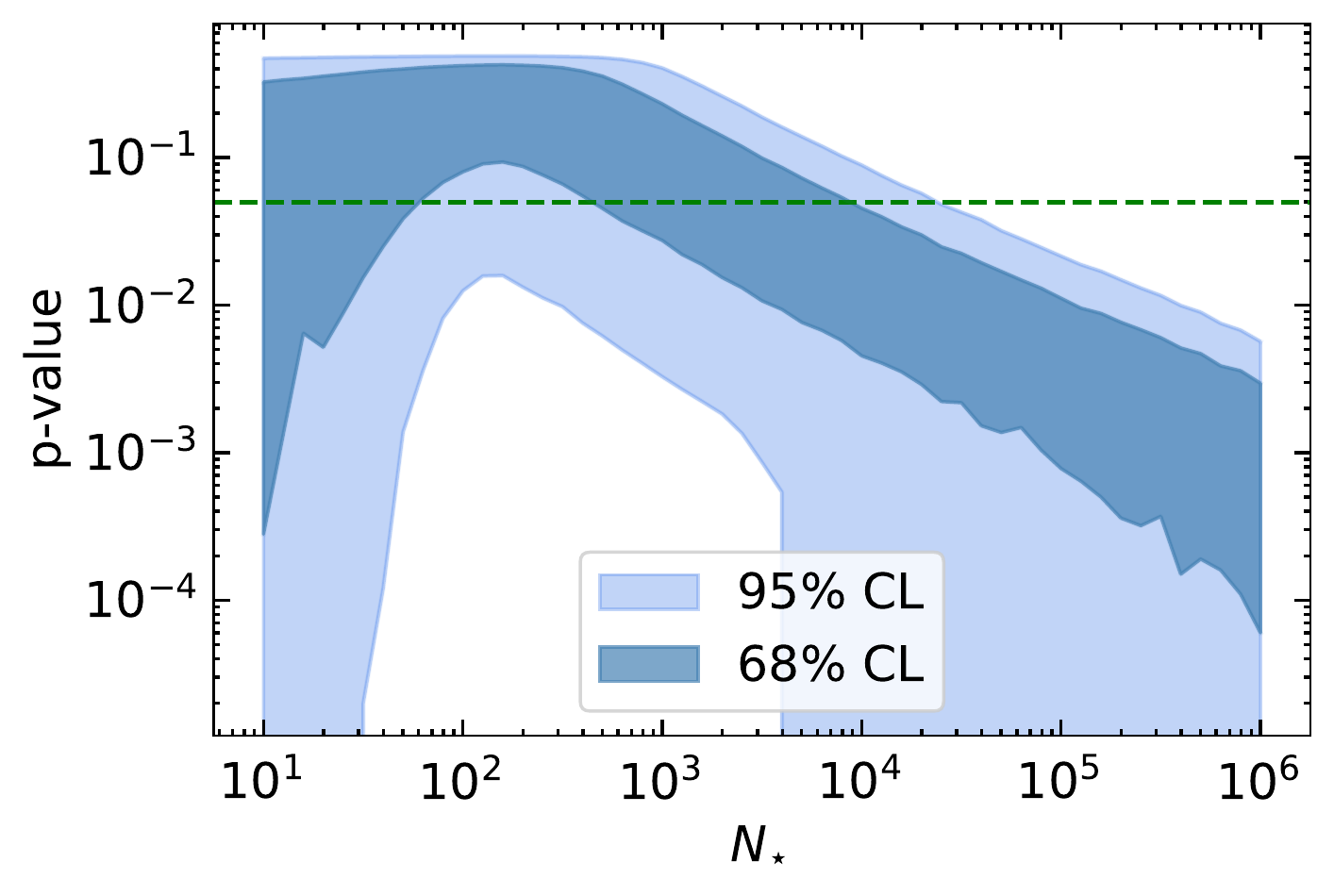}
  \hspace{1cm}
  \includegraphics[width=.45\textwidth]{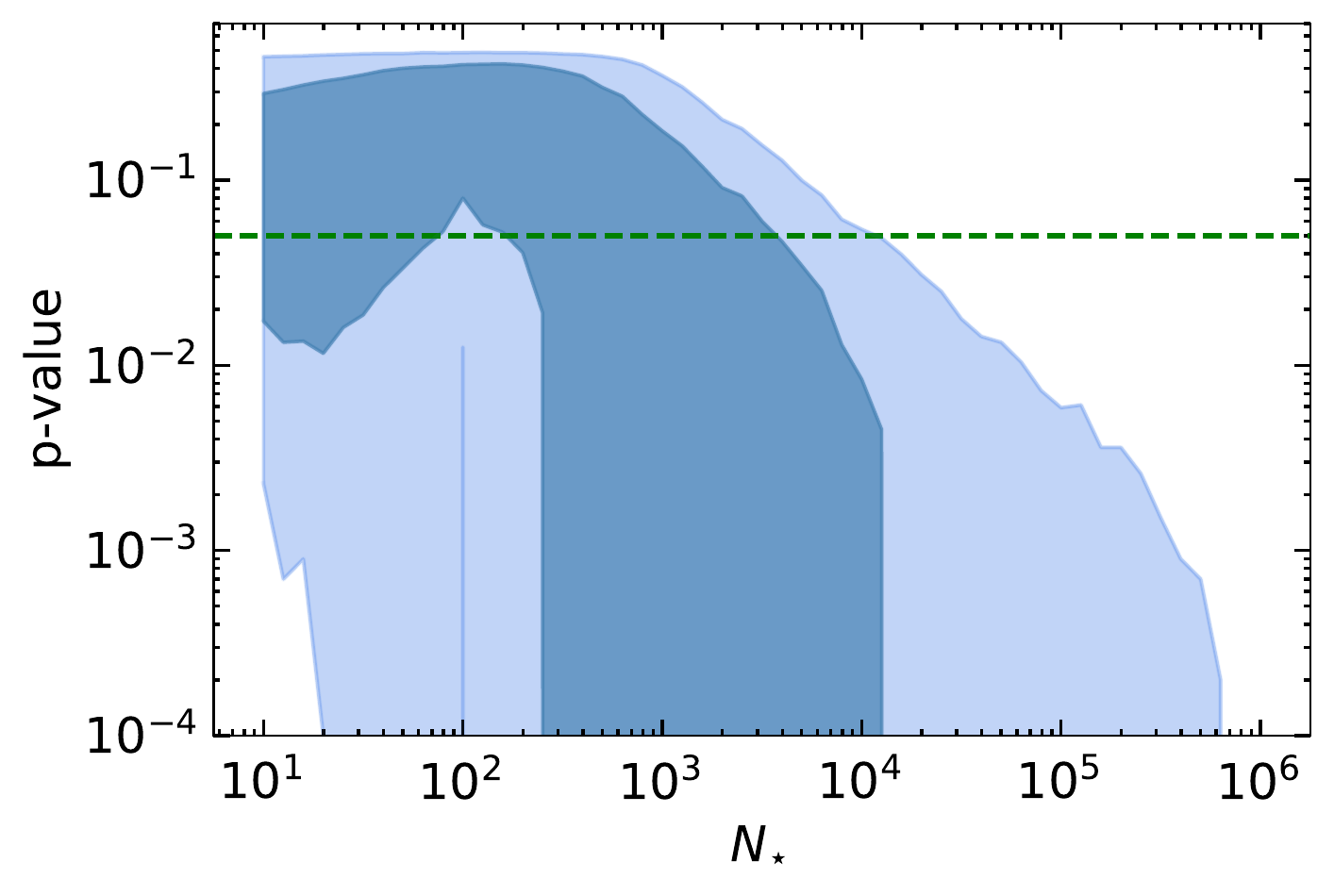}
  \caption{$p$-value results using expected future neutrino data with ten years of IceCube-Gen2 (left panels) and KM3NeT exposure (right panels) by assuming from top to bottom $N_{s} = \infty$, $N_{\star}=10^{4}$ and $10^{2}$.}
  \label{fig:pval_future}
\end{figure}

The same approach is applied by using ten years of KM3NeT exposure, from which the results are presented in the right panels of Fig.~\ref{fig:pval_future}. The top right panel shows the case when assuming an isotropic neutrino sky, and we find that sources below $N_{\star} = 10^{4}$--$3\times 10^{4}$ can be excluded at 95\% CL, where the range represents the $1\sigma$ band. The cases by assuming $N_{\star} = 10^{2}$ and $10^{4}$ are shown in the middle and bottom right panels of Fig.~\ref{fig:pval_future}, respectively. 
The $N_{\star} = 10^{2}$ situation shows a lower exclusion band of $N_{\star} = 10$--$80$ and an upper exclusion band of $N_{\star} = 2\times10^{2}$--$4\times10^{3}$. Whereas assuming $N_{\star} = 10^{4}$ results in a lower exclusion band of $N_{\star} = 4\times10^{2}$--$5\times10^{3}$ and an upper exclusion limit above $N_{\star} = 3\times10^{4}$.
Comparing the results of IceCube-Gen2 and KM3NeT we find similar exclusion trends, possibly due to the factor 10 larger effective area of IceCube-Gen2 while a factor 10 improved angular resolution of KM3NeT. 
Therefore, these two future detectors, IceCube-Gen2 and KM3NeT, have similar sensitivities from the APS measurements, confirming theoretical finding of Ref.~\cite{Ando:2017xcb}.

\subsubsection{Case of fractional contributions}
\begin{figure}[h!]
	\centering
	\includegraphics[width=0.6\textwidth]{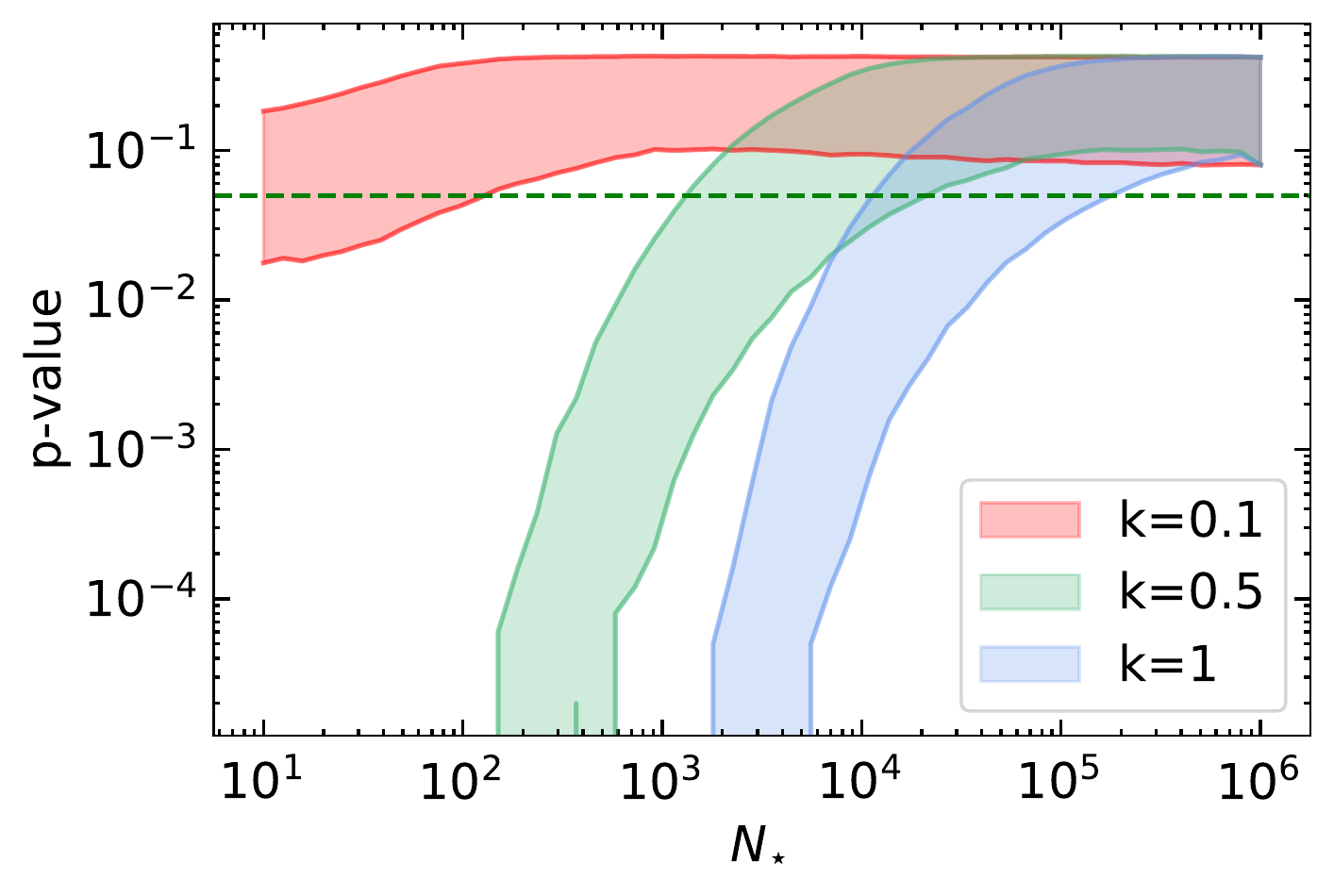}
	\caption{$p$-value results with 10 years of IceCube-Gen2 exposure, showing different fractions to the isotropic neutrino intensity: $k=0.1$ (red), $k=0.5$ (green) and $k=1$ (blue), and the $95\%$ exclusion limit (green dashed line).}
	\label{fig:pval_fr} 
\end{figure}

Up to now, we assumed that a single population of the sources characterized with $N_{\star}$ fully contributes to the measured neutrino intensity. However, this might not be the case and different source populations can dominate the neutrino sky. We therefore also consider smaller contributions with respect to the isotropic intensity. In particular, we define the total intensity as $I_{\rm{tot}}= k I_{N_{\star}} + (1-k) I_{\rm iso}$, where $I_{N_\star}$ is the intensity due to $N_\star$ population and $I_{\rm iso}$ is that from the isotropic component ($N_s = \infty$), and study the exclusion patterns for the fractions $k=0.1$ and $k=0.5$. 
We perform $5\times10^{4}$ Monte Carlo simulations using 10 years of IceCube-Gen2 exposure. Figure~\ref{fig:pval_fr} shows the results with $k=0.1$ (blue), $k=0.5$ (green) and $k=1$ (red). We find no significant exclusion in the case of $k=0.1$, which is consistent with previous work that predict a blazar contribution of up to $\sim 10\%$. In the case of $k=0.5$, we do find a stringent exclusion limit of $N_{\star} = 10^3$--$2\times10^4$, where the range represents the $1\sigma$ band. Since the results of IceCube-Gen2 and KM3NeT show similar exclusion trends (Fig.~\ref{fig:pval_future}), we also expect similar results for the latter.

\subsubsection{Varying the angular resolution}
\begin{figure}[h!]
	\centering
	\includegraphics[width=0.6\textwidth]{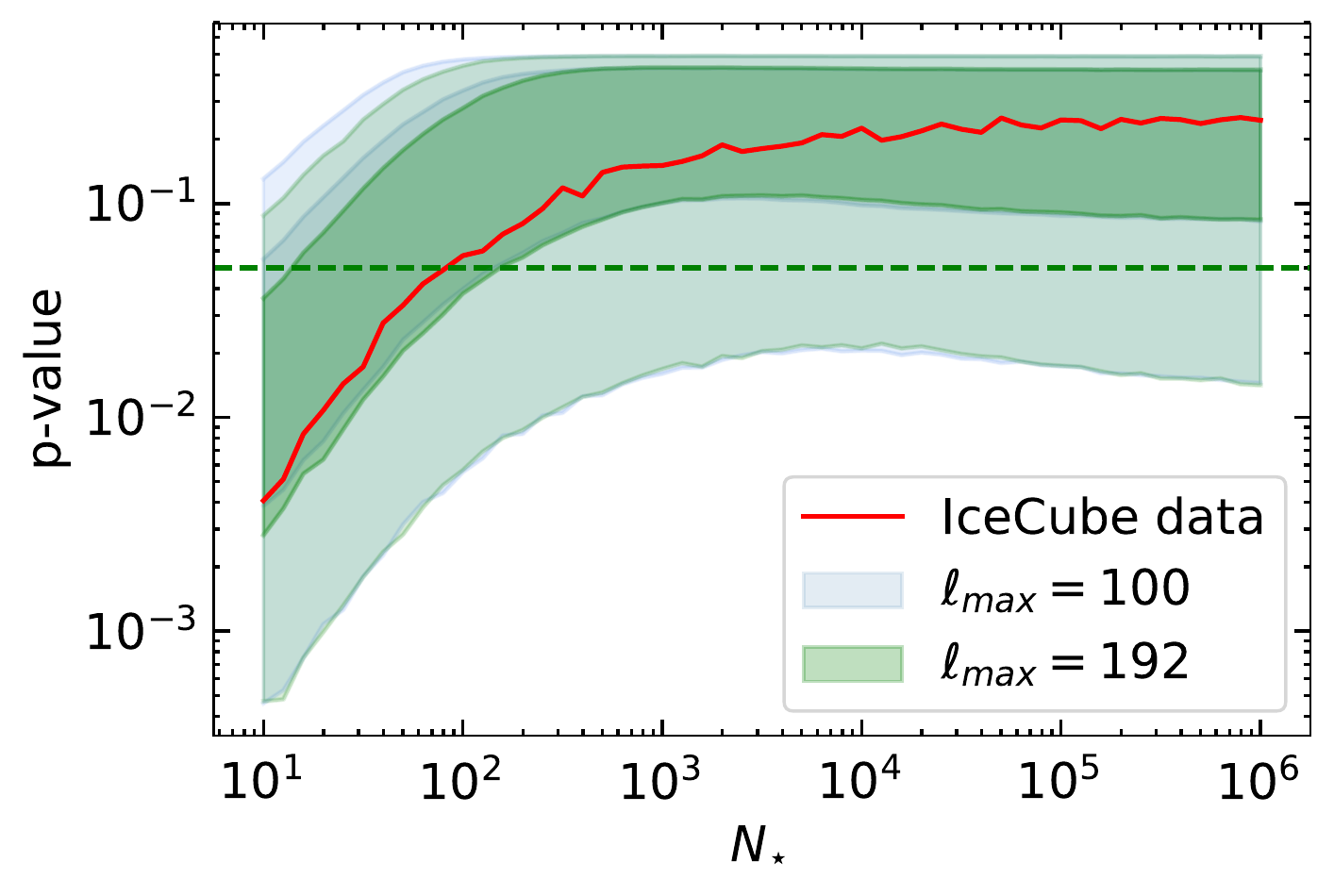}
	\caption{$p$-value result for 2 years IceCube using $\ell_{max}=100$ ($\sigma=1^{\circ}$, blue) and $\ell_{max}=192$ ($\sigma=0.5^{\circ}$, green). }
	\label{fig:pval_lmax} 
\end{figure}

We checked if our results would change by assuming weaker angular resolutions for IceCube, IceCube-Gen2 and KM3NeT, and applied the method using half of the angular resolution. Figure~\ref{fig:pval_lmax} shows the $p$-value result in the case of two years of IceCube exposure with $\ell_{max}=100$ ($\sigma=1^{\circ}$, blue) and $\ell_{max}=192$ ($\sigma=0.5^{\circ}$, green), and we find no significant change in result.

\section{Discussion}\label{sec:4}
BL Lacs, flat-spectrum radio quasars (FSRQs), radio galaxies, and starburst galaxies are generally thought to be the main contributers of the measured intensity~\cite{Aartsen:2016lir,Loeb:2006tw,TAM,Hooper2016b,MuraseWaxman}, and we therefore relate the obtained exclusion values to these source classes. Reference~\cite{Ando:2017xcb} found the ($\alpha, \beta, N_{\star}$) parameterization for BL Lacs, and starburst galaxies of ($2.5, 1.7, 6\times 10^{2})$ and ($2.5, 1.0, 10^{7}$), respectively, and it is also argued that the typical $N_\star$ value for radio galaxies is $\sim$10$^5$. The parameterizations are obtained from gamma-ray observations by assuming a linear relation between the gamma-ray and neutrino luminosity, $L_{\gamma} \propto L_{\nu}$, which can be estimated if neutrinos are produced through hadronic interactions. 
Lepto-hadronic sources show a more complicated luminosity relation, e.g., $L_{\gamma} \propto L_{\nu}^2$ for BL Lacs~\cite{Tavecchio2015} or $L_\gamma \propto L_\nu^{1.5}$ for FSRQs~\cite{MID}. We thus note that a more realistic parameterization for lepto-hadronic source classes could give stronger constraints. In fact, Ref.~\cite{Krauss:2018tpa} mentions that blazars produce high-energy neutrinos for less than $56.2\%$ through hadronic interactions. 

Comparing $N_{\star}$ from the aforementioned source classes with the exclusion limits obtained from analyzing the APS using two years of IceCube exposure, we find that the limits are not significant yet. Increasing the exposures to 10 years for the neutrino detectors IceCube, IceCube-Gen2 and KM3NeT, shows more promising results. If we assume an isotropic sky in the future, shown in the top panels of Fig.~\ref{fig:pval_future}, we find that BL Lacs are excluded as dominant contributers to the intensity with $95\%$ CL. The non-detection of clustered events does not constrain source classes with large $N_{\star}$, which produce neutrinos typically in cosmic ray reservoir models like starburst galaxies. 
However, if, in the future we observe clustered events, we show that our analysis can constrain large $N_{\star}$ populations as well. Indeed, the middle and bottom panels in Fig.~\ref{fig:pval_future} show respectively the cases assuming a $N_{\star} = 10^{4}$ and $10^{2}$ neutrino sky. The contribution of starburst galaxies ($N_\star \approx 10^7$) is significantly constrained in both cases, and radio galaxies ($N_\star \approx 10^5$) are well constrained in the case of a neutrino sky with $N_\star = 10^2$.

\begin{figure}[h!]
	\centering
	\includegraphics[width=0.8\textwidth]{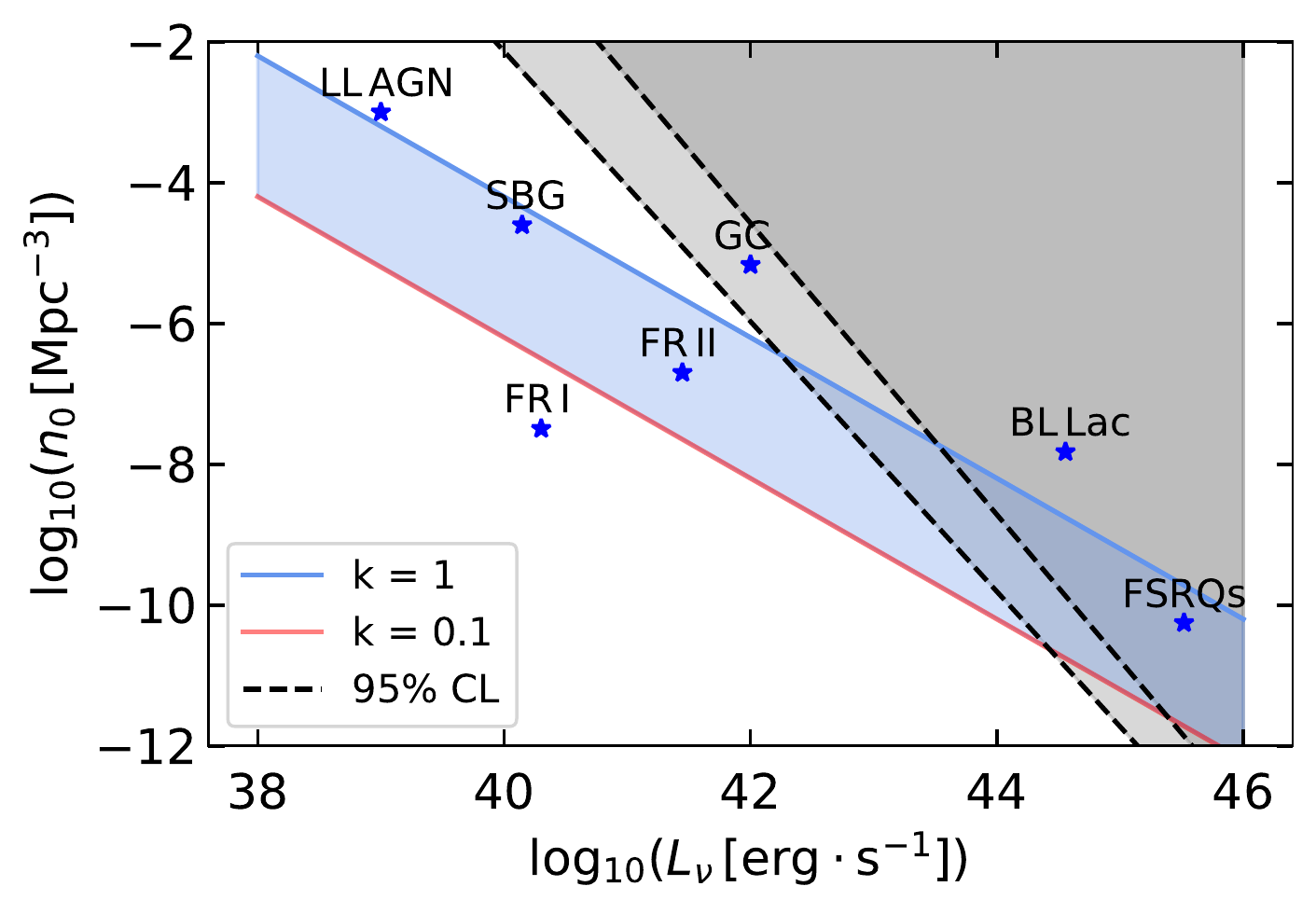}
	\caption{Exclusion region with 10 years of IceCube-Gen2, shown for the neutrino luminosity, $L_{\nu}$, against the local number density of neutrino sources, $n_{0}$. The gray exclusion region is obtained by assuming an isotropic neutrino sky in future, where the two black dashed lines are the $95\%$ exclusion limits, and all sources lying in that region are thus excluded. The blue region represents the observed diffuse neutrino emission, taken from Ref.~\cite{Mertsch:2016hcd}, where the neutrino source emission contributes for $k=1$ (blue) and $k=0.1$ (red). }
	\label{fig:lum} 
\end{figure}

In order to compare the results with more physical values, we convert $N_{\star}$ to the luminosity using Eq.~(23) of Ref.~\cite{Ando:2017xcb}, assuming that all sources have equal luminosity. We use the relation between the luminosity and the local number density of neutrino sources from Ref.~\cite{Mertsch:2016hcd}, which is derived from the observed diffuse flux with a $\propto E_{\nu}^{-2}$ neutrino spectrum, and by assuming no redshift evolution. The result is shown in Fig.~\ref{fig:lum}. The blue band represents the region that explains the observed diffuse flux,  where the upper limit (blue) represents a contribution of $k=1$ to the diffuse flux, and the lower limit (red) a contribution of $k=0.1$. The black dashed lines show the $95\%$ exclusion limits, taken from the results in Fig.~\ref{fig:pval_fr}, whereby the width represents the $1\sigma$ band. The gray region is excluded when no clustering of events is found in future. 
BL Lacs lie in the exclusion region, and moreover, lie outside the blue band representing the diffuse flux. Also galaxy clusters (GC) lie above the blue band, and inside the $1\sigma$ exclusion band, and are thus too bright to explain well the isotropic neutrino emission. FSRQs lie inside the diffuse flux band, however, due to their small number density and large luminosity, a significant clustering of events is expected, and they are therefore testable sources with APS. Low-luminosity AGNs (LL AGN), Fanaroff-Riley galaxies of type II (FR-II) and starburst galaxies (SBG) are consistent with the isotropic neutrino emission. These sources could thus be observed as point-sources in the near future with IceCube-Gen2 and KM3NeT. FR-I will be difficult to observe due to their low luminosity and number density, and they will hardly contribute to the high-energy neutrino emission. 

Up to now we assumed no source evolution, which is not the case for all source populations. Most BL Lacs have a positive evolution, except for the low-luminosity, high-synchrotron peaked BL Lacs which have a negative evolution~\cite{Ajello:2013lka}. Choosing positive (negative) redshift evolution will shift the blue band in Fig.~\ref{fig:lum} to lower (higher) normalization. For example, using positive evolution following the star-formation rate in the Universe yields a factor $\sim$3 times smaller normalization~\cite{Mertsch:2016hcd}. This will not impact our exclusion values based on anisotropies, but will impact whether source populations are consistent with the diffuse flux band, as shown in Ref.~\cite{Mertsch:2016hcd}. 

We tested that our results are robust against varying the parameters $F_{0}$, $\alpha$ and $\beta$. However, the results will probably change by varying the flux normalization and spectral index, for which we applied $\gamma = 2.2$; see Eq.~(\ref{eq:spectra}). A spectral hardening at high energies has been suggested~\cite{kowalski2017}, and since the spectral index depends on the neutrino production model, this could indicate multiple production mechanisms or source populations~\cite{kowalski2017}. If the spectral index would harden with better statistics in the future, we expect to have stronger constraints.

\section{Conclusion}\label{sec:5}
We have derived constraints on $N_{\star}$, by analyzing the angular power spectrum for the current and future high-energy up-going muon neutrino data, detectable by IceCube, IceCube-Gen2 and KM3NeT. The constraints were established by applying statistical distributions to the flux of the individual extragalactic sources. 
Our finding, by using two years of IceCube observations~\cite{Aartsen:2015rwa} with only 21 high-energy up-going muon events, is already a constraint on the characteristic source number $N_{\star}<82$ with $95\%$ CL, excluding the dominant contribution from very bright source populations.
Comparing this result with known source classes is not effective yet, however, with 10 years of future neutrino data using IceCube, IceCube-Gen2 and KM3NeT exposure, we expect to significantly constrain BL Lacs and FSRQs if the distribution remains isotropic.
In particular, 10 years of IceCube-Gen2 observations excludes with $95\%$ CL characteristic source number less than $10^{4}$--$2\times10^{5}$, and KM3NeT less than $10^{4}$--$3\times10^{4}$, where the range represents the $1\sigma$ band and where BL Lacs are found to have $N_{\star}=6\times10^{2}$.
On the other hand, by observing bright sources in the future we can also find constraints on weak source classes with large number densities, such as starburst galaxies ($N_{\star} = 10^{7}$), which could still be the case with current isotropic measurements.
The angular power spectrum analysis on future neutrino data has been found to be a powerful probe to understand what astrophysical sources are dominating the neutrino sky, and in particular to predict what source classes will be observable with future neutrino telescopes, illustrated in Fig.~\ref{fig:lum} for various sources.

\acknowledgments
We thank Aart Heijboer for helpful discussions, especially for specification of KM3NeT. SA acknowledges support by JSPS KAKENHI Grant Numbers JP17H04836, JP18H04340, and JP18H04578.

\bibliographystyle{utphys}
\bibliography{refs}

\end{document}